# An ecology of participation for fusion energy development

Aditi Verma[1], Katie Snyder, Andrea Morales Coto, Nathan Kawamoto, Daniel Hoover, Ana Kova, Sara Eskandari, Stephanie O'Malley, Jared Owens, Mahmud Farooque, Gabrielle Hoelzle, Stephanie Diem, Kevin Daley

**Abstract**

Decisions that will shape future fusion facilities — production of waste, management of tritium, achievement of safety, and impacts on land, water, and local communities — are increasingly becoming sociotechnical in nature. For prior energy technologies, fission among them, such decisions were made top-down and met sustained public opposition. Research shows such opposition is rarely a deficit of public understanding and is instead rooted in place-specific environmental, health, sociocultural, and economic concerns and in distrust of how technologies are selected and sited. We argue that fusion developers have an opportunity to design differently, and we propose an ecology of participation: multiple, complementary modalities of engagement sustained over time, replacing the one-off public consultations now typical of large infrastructure projects. We synthesize human- and environment-centered design frameworks and introduce a framework that reinterprets opposition as a divergence between communities and technology developers in values, norms, or design characteristics. We present six case studies from our work: participatory technology assessment focus groups, participatory design workshops, immersive virtual-reality models of a fusion facility, the Global Fusion Forum platform, the Imaginary Energies speculative design platform, and Heartbeat, an arts-based sound installation. Mapping these onto the values-norms-design framework shows how each interrogates or closes a different part of the expert-public gap, and exposes two limitations of fusion public engagement efforts writ large: developers are seldom asked to articulate their own values and norms, and the environment is represented only indirectly. We close with five guiding questions for fusion researchers building their own participatory engagements.

[1] aditive@umich.edu Assistant Professor of Nuclear Engineering and Radiological Sciences, University of Michigan.

Author affiliations: Aditi Verma (Department of Nuclear Engineering and Radiological Sciences, University of Michigan), Katie Snyder (Program in Technical Communication, University of Michigan), Andrea Morales Coto (Fastest Path to Zero Initiative, University of Michigan), Nathan Kawamoto (University of Michigan), Daniel Hoover (University of Michigan), Ana Kova (Science Illustrator), Sara Eskandari (Visualization Studio, University of Michigan), Stephanie O'Malley (Visualization Studio, University of Michigan), Jared Owens (Design Science Program, University of Michigan), Mahmud Farooque (Consortium for Science, Policy and Outcomes, Arizona State University), Gabrielle Hoelzle (Fastest Path to Zero Initiative, University of Michigan), Stephanie Diem (University of Wisconsin–Madison), Kevin Daley (Fastest Path to Zero Initiative, University of Michigan)

# Introduction

As fusion researchers, technology developers, and policymakers plan for fusion energy commercialization and potential widespread use by mid-century, it is becoming increasingly important to consider the potential socioeconomic, environmental, and geopolitical impacts of fusion energy systems (*1–4*). For the last several decades, efforts directed towards the realization of fusion energy have been focused on the grand scientific challenges of creating the conditions for fusion to occur – from the long-term confinement of plasmas and reliable plasma diagnostic systems to suppression of disruptions and most recently, the achievement of a burning plasma. Recently, with the release of the National Academies of Science, Engineering, and Medicine's (NASEM) report on commercial fusion energy (*5*), researchers are now tackling the scientific and engineering challenges of commercial fusion energy – from the development of novel materials that can withstand the harsh conditions of a fusion device to harnessing fusion energy. As these grand challenges of fusion gradually find resolution, fusion developers are shifting attention to creating demonstration and even powerplant-scale fusion systems – transitioning the field of fusion from being focused on science to a growing focus on engineering. Engineering decisions being made throughout the fusion energy system design processes are not just technical but, in fact, *sociotechnical* in nature. They will shape how much and what kinds of waste fusion energy systems will produce; how much tritium will circulate through these systems, how it will be managed and the potential environmental and health impacts of tritium handling to workers and nearby communities; the levels of safety fusion energy systems will achieve through inherent design, automation, or human intervention; the levels and kinds of employment opportunities fusion plants will create; the size of emergency planning and exclusion zones around fusion plants, if any; the ability of fusion systems to integrate well with other energy sources on the grid; and the land and water use of fusion energy systems (*6*). Even this list of the impending sociotechnical decisions concerning the design of fusion energy systems and their potential societal and environmental impacts is likely incomplete and will grow along with our development, use, and understanding of fusion energy systems. Certainly, fusion developers are recognizing the importance of resolving challenges, as evidenced by the creation of many public-private partnerships in the United States and abroad (*7*) to build toward sociotechnical readiness of fusion energy technologies (*8*, *9*).

For many energy technologies, these sociotechnical decisions have been made in a relatively hierarchical manner by the designers and developers of technologies, with those experiencing the societal and environmental impacts of technologies having little say in how these decisions are made.The development of nuclear fission energy systems for grid-scale deployment is one example. Beginning with the Shippingport Atomic Power Station in 1957, this project was the result of the US Government's Atoms for Peace Program, and its location was largely determined by desirable geography and a utility company bidding competition (*10*). The power station was completed quite quickly, without community consultation or meaningful consideration of the near- and long-term environmental impacts (*11*). This top-down approach was applied widely by early nuclear power designers and ultimately led to significant public opposition, unsolvable controversies and, in some cases, moratoria on siting and design (*12*). These challenges to nuclear energy technology adoption, in our view, were largely driven by a failure of design imagination.

Here, we contend that fusion energy developers have a unique opportunity to make considered, deliberative, and inclusive design choices, such that fusion energy technologies are well-suited to adaptable, widespread adoption. Toward that end, we argue for approaching design as an 'ecology of participation' – we advocate for understanding energy systems design as an ecology of systems and time, including, past, present, and future realities, in the contexts of sociotechnical concerns and commitments to environmental thriving. More specifically, we articulate here a human- and environment-centered approach to designing fusion energy systems where there are no (or few) externalities. In this paper, we begin by describing human and environmental concerns and constraints to energy system technology adoption. We then synthesize insights from technology design frameworks that bring human and environment-centered perspectives into the technology design process. We present several case studies highlighting how these human and environmental-centered perspectives can inform fusion energy development. In closing, we map our case studies onto a values, norms, and design divergence framework, reflect on our own ecology of participation, and offer a set of guiding questions to fusion researchers seeking to build participatory engagements of their own.

# Why publics oppose clean energy technologies

In many fields of engineering, the rejection of technologies or industrialization is attributed to the deficit model (*13*)– that opposition to a technology is the result of an incomplete understanding of the benefits of a technology, and that concerted public education efforts can quell concerns and earn deserved social license. In reality, the causes of opposition to a technology, in this case energy technologies, are much more complex, diverse, and context-specific, and policy and engagement efforts rooted in the deficit model explanation are likely to fail (11-28).

A growing body of scholarly work examines the sources of opposition to energy technologies, including clean energy technologies and renewables. This work finds that energy technology opposition, far from stemming from a lack of understanding, is instead rooted in location-specific concerns about environmental degradation, aesthetic impacts, health and safety concerns, concerns about socio-cultural disruption, concerns about economic impacts, as well as concerns about planning, procedural, and governance issues, as summarized in greater detail in Table 1.

Table 1. Sources of opposition to energy projects. Importantly, these categories are not mutually exclusive. For example, environmental concerns may also be health and safety concerns or have economic or sociocultural impacts. In other words, sources of opposition are multiple, overlapping, and interconnected.

| | |
|---|---|
| Environmental and landscape concerns | ➔ **Visual impacts and aesthetic impacts**, which include concerns about the industrialization of natural landscapes, or the spoiling of scenery as a result of the construction of large structures like cooling towers, wind turbines, or solar farms (*14–20*).<br>➔ **Damage to local ecosystems**, including rare species of plants and animals, resulting from the energy facility itself or the construction of transmission infrastructure (*16*, *21*)<br>➔ **Pollution risks**, including noise pollution from energy facilities, in particular from wind turbines and even high voltage transmission lines (*20*), thermal pollution from cooling towers, as well as chemical run-off from solar panels (*14*, *22*) |
| | ➔ **Water-related impacts**, including impacts on local water supply, rainwater drainage, and alterations in natural water flows, particularly from hydropower (*20*, *22*)<br>➔ **Improper use of land**, which may include the conversion of prime agricultural land, leading to a loss in food production, the alteration of unique geological formations, and long-term risks to the land even after project decommissioning (*17*, *19*, *20*) |
| Health and safety concerns | ➔ **Risks of accidents**, particularly for nuclear power plants (*23*)<br>➔ **Chemical and radiotoxicity of materials** for both solar and nuclear during normal operation, accidents, and waste management (*14*, *17*)<br>➔ **Ongoing impacts to human health and quality of life**, such as stress, nausea, and sleep deprivation, caused by noise pollution from wind turbines (*21*, *24*) |
| Sociocultural concerns | ➔ **Threats to traditional ways of life** and livelihoods, including the right to self-determination, particularly in indigenous communities (*20*, *25*). Disruption of local customs, community cohesion is also a concern (*20*).<br>➔ **Disruption to place attachment** and identities due to strong emotional bonds formed with landscapes or neighborhoods, which may irreversibly be altered by new infrastructure (*16*, *20*, *26*)<br>➔ **Threats to cultural heritage sites,** including culturally significant landscapes (*20*, *21*) |
| Economic and distributive justice concerns | ➔ **Negative impacts on property values** (*16*, *17*)<br>➔ **Inadequate financial compensation** for the host community (*20*, *27*, *28*)<br>➔ **High costs for local governments** (*22*)<br>➔ **Exploitative business practice**s from energy companies (*19*)<br>➔ **Loss of revenue from tourism** (*29*) |
| Planning, procedural, and governance concerns | ➔ **Insufficient public participation and consultation** in the decision-making process, including opaque, top-down decision-making processes such as "decide-announce-defend" (DAD) (*17*, *20*)<br>➔ **Perceived and real unfairness** and lack of procedural justice in siting energy projects, including efforts to limit local involvement and curb opposition (*20*, *22*, *26*, *30*)<br>➔ **Distrust** in electric utilities, project developers, government authorities, and regulators due to perceived self-interest, deceitful tactics, inconsistent or unclear regulations, or failure to protect community interests, and misalignment of values (*17*, *18*)<br>➔ **Lack of community ownership** of a project (*31*) |

In summary, communities resist energy technologies either as a result of material concerns about the local impacts of those technologies or because of a loss of trust in decision-makers and the governance processes through which new technology projects are selected, developed, and sited. The causes of opposition are not universal for a particular technology and are mediated by local culture, history (including industrial history), and geography. For these reasons, direct interaction with potential host communities is key to the successful development of energy projects.

The implications of these findings are two-fold for new energy technologies and projects broadly and fusion energy specifically: (1) successful context-appropriate adoption of energy technologies calls for giving local communities a say in key design and planning decisions; and as a corollary, (2) more deliberative and participatory decision-making processes are needed. The sections below provide a map of human- and environment-centered design practices, from which we will draw pieces to build the ecology of participation we envision.

# Bringing human and environment-centered perspectives to fusion technology design

A variety of design strategies have been developed as a way to bring human- and environment-centered perspectives into design practices, broadly construed. From user-centered design to participatory design and design justice, these strategies seek to improve outcomes for users as an overarching goal, but some also aim to mitigate long-running inequities, injustices, or other imbalances of power that traditional design strategies reify or otherwise ignore. Notably, energy systems designers have not historically identified the people who "use" energy produced by said systems as "users." The terms "stakeholder," and more recently, "rightsholder," (*32*) are more typical ways to conceptualize this group. But by any definition, people and communities are *users* of energy systems infrastructure, and a growing body of research suggests that their experiences, knowledge, and perspectives should be valued, solicited, and included in shaping energy systems design (*33*), (*34*), (*35*). The sections below outline a series of design strategies meant to center humans and the environment in design practices and protocols as a way to trace evolution of design theory and argue for an ecologized participatory approach (*36*) to energy systems design.

The theory and practice of bringing human-centered perspectives into the design process can be traced back at least two thousand years to Ancient Greece and the idea of *techne*, translated as 'craft.' Techne means *practical knowledge* and, specifically, knowledge of how to make something very well-suited to its users and their needs. It also implies contextual knowledge about the people, materials, and methods affected by or involved in the design process and its resulting products (*37*). The English word "technology," of course, emerged from this origin, but the idea that *technology* should be inherently user-centered did not endure. As noted above, much technological innovation and design has taken place over the ensuing millennia that did not consider the values, needs, and contexts of its intended users. More recently, however, at the turn of the 20th century, Scandinavian scholars began exploring the idea of "use" in design contexts, believing that the prioritization of *functionality* in design could "reform" or improve everyday life. By the 1960s and 1970s, Scandinavian designers were developing *user-centered design* methods meant to observe users in specific contexts. They then integrated these observations into their ideation processes, and also began tying design objectives to human-centered concerns like equity and accessibility (*38*).

The shift from *user-centered* to *human-centered* design (HCD) was driven in part by an interest in seeing users as whole people immersed in meaningful socio-cultural contexts (*39*). While human physical or physiological realities remain central to this approach, an interest in using design to improve human work experiences, with increasingly liberatory aims, became a motivating factor. Importantly, human-centered design was largely developed as a way to create better software and other digital tools for people in work contexts. The ambition here was to establish a technology design framework where new products and processes serve humans and improve their overall quality of life rather than simply solving decontextualized, technology-centered problems (*39*). Key questions about whether emerging technologies might deskill humans, suppress creativity, or limit human potential were also important in establishing this framework; of course, these questions have become increasingly important with the emergence of artificial intelligence (AI) technologies.

Philosophies of technology have informed HCD as well, including the idea that technology shapes social life and expectations, even as social expectations and cultural contexts may influence technological development (*40*). Some argue that technological development plays too significant a role in shaping social life, and argue for political interventions or slowing the pace of innovation to allow time for assessment, critique, or change in trajectory (*41*, *42*). As a recent and specific example, some have argued for slowing the pace of AI development to allow time to develop appropriate guardrails and assess whether emerging technologies are truly human-centered and designed to serve a shared common good (*43*). In nuclear technology contexts, human-centered design has been applied to energy system project management (*44*) and, more typically, to nuclear control room design (*45*), (*46*). These examples tend to focus on the physiological or cognitive needs of humans in the design context rather than philosophical concerns stated above.

Critiques of human-centered design suggest that the process is not realistic or results in overcomplicated workflows and unhelpful products (*39*, *47*). More recently, some critics suggest that HCD is not equipped to grapple with the complexities of emerging digital technologies because of its close alignment with profit-driven business models and inattention to environmental sustainability (*48*).

## Value sensitive design

Value sensitive design (VSD) is a method for examining or unpacking the ways corporate, cultural, or individual design values shape design requirements. At the same time, VSD offers a strategy for inviting users and communities to share their values with designers, and for designers to use those values to create meaningful and useful design requirements. This process is often interdisciplinary in nature, requiring technical experts to translate particular values into norms. In other words, if one values environmental sustainability, experts in ecology or the environment can explain what the features or qualities of sustainability might be or mean in a given design context. Those qualities are translated into norms, and those norms then shape the development of specific design requirements, along with the means to meet, achieve, or preserve established norms (*49*). Of course, the translation of values into design requirements is also shaped by existing legislation, codes, and standards. In some cases, values translation may reshape these codes – in other cases, trade-offs may be necessary given these and other constraints like financial viability and market demand. VSD does not guarantee that users' values will net the exact outcome they desire, but the process offers a pathway for communication and to build trust and mutual understanding between users and technology designers. It also offers a means to grow designers' moral imagination, or their ability to imagine the ways human values might shape or be integrated into design practices and processes (*50*). In an ideal context, this process of communicating values would be ongoing across the lifecycle of a long-lived project like a fusion energy system, particularly as the system manages waste and moves to decommissioning and long-term storage. Since its emergence in the early 1990s in the context of human-computer interaction, this practice has been applied to a wide range of disciplines and design spaces, from animal welfare (*49*) to autonomous weapons (*51*), smart metering systems (*52*), and healthcare (*53*).

Notably, VSD comes with some criticism or limitations. For example, establishing an equitable and transparent method for selecting and weighting community values in a given design context can be quite challenging. In recent years, an increasing number of researchers advocate for the "best worst method" (BWM) as a promising approach, particularly when trying to rank conflicting values. Notably, this method relies on expert opinion. In some cases, for example, values may be drawn from research literature rather than directly from users and communities as a matter or primary research. Some researchers have expressed concerns that this method might overlook or misrepresent individual user value expressions and preferences (*52*). It is also worth noting that in a design process, it may not be possible to translate all values into design characteristics, and some values may lead to potentially conflicting design choices.

## Participatory design and co-design

Participatory design (PD) has its origins in 1970s Scandinavia as a method for designing with users, infusing user values into the design process, and approaching conflict or contradictions as valuable resources for improving design outcomes (*54*). The idea is founded on a democratic belief that users should have a voice in shaping the products and processes with which they engage. Further, this approach acknowledges that individuals and communities have tacit knowledge or lived experience that can be as valuable as expert knowledge when applied to technological design processes (*55*). In practice, participatory methods are somewhat variable, ranging from community workshops to public forums, focus groups, and individual interviews. Participation in "living labs," or spaces where design work is taking place in real time and real-world contexts, is also increasingly common as a method for inviting users into the participatory design process (*56*).
Of course, challenges of PD design are also important to note, including establishing accurate methods of assessment (how do we know this approach "worked," and what does "worked" mean in each design case?), securing sustainability over a long-term project, mitigating knowledge hierarchies, and establishing meaningful engagement across the course of a project. Notably, PD-informed projects are almost uniformly initiated by those with technological expertise wishing to bring a new product or process to market. Communities or users are not typically in a position where they can, or have space to consider, inviting the creation of a new product or project that addresses their unmet needs. At the same time, in the context of fusion energy technologies, we know that many communities are asking for more efficient, sustainable, and cost-effective energy infrastructure (*57*). (Perhaps, as public discourse about fusion energy increases and grid-scale fusion energy becomes viable, communities could be empowered to initiate or co-lead on fusion energy infrastructure design in their local contexts.) This approach moves more towards a co-design model, and could perhaps shift long-standing power imbalances and perhaps speed the process of fusion energy power systems development and integration.

Co-design, then, is a strategy that seeks to mitigate imbalances in power sometimes inherent in traditional design methods by inviting users to be partners in design. In other words, rather than simply gathering participants' ideas, perspectives, and knowledges to inform technological design, co-design moves a step further and gives participants an active role in the design process across the project timeline (*58*). Like VSD, co-design practices can invite open and consistent communication between users and designers, but here the conversation is not only focused on sharing and translating values. The work is collaborative and hands-on, with a focus on creativity. The front-end of this design process can take longer than traditional approaches, but is particularly desirable for the purposes of creating an excellent product or process, suited to specific user needs and contexts. Importantly, the co-design process, in some cases, has been initiated by communities rather than by experts (*59*). In these cases, communities identify a problem or design challenge they want to solve, and then invite relevant experts to collaborate with them in design and implementation. Co-design is also sometimes used synonymously with the term "co-creation," and emphasizes values like environmental sustainability, interdisciplinarity, and even spirituality (*60*). In sum, it is rooted in PD and HCD, but offers different opportunities for democratizing the design process and empowering users to initiate or take the lead in the design space.

## Participatory technology assessments

A participatory technology assessment (pTA) can be viewed as a precursor or parallel activity to participatory design. It refers to a group of methods and systematic procedures to involve "societal groups–stakeholders, affected citizens, non-experts, and the public in general–in assessing technology and its consequences" (*61*). A central aim of pTA is to involve non-experts who are representative of the general population, but who, unlike organized interests such as professional societies, industry organizations, and advocacy groups, are generally not represented in science and technology discourse and decision-making. Such assessments, developed in coordination with subject matter experts, challenge the traditional expert-centric technology assessment paradigm, in which users and those potentially impacted by a new technology and its social consequences have had no say in its assessment. As with participatory design, there is no single participatory technology assessment method; rather, there is a still-evolving collection of methods (*62*) that have shared normative and procedural underpinnings.
Participatory technology assessments emerged in the 1980s in Western Europe, prominently in Denmark and the Netherlands, as a response to growing demands for peace, sustainability, and more deliberative democracy placed on their technology assessment (TA) agencies (*62*),(*61*). Given adjacency to the deliberative democracy movement, and its central aim to add public participation ('p') to expert assessment ('TA'), pTA embraced many deliberative democracy methodologies such as citizen assemblies (*63*), planning cells (*64*), citizen juries (*65*), and deliberative polling (*66*) to engage citizens. As such, the first generation of pTA could be viewed as an application of deliberative democracy practices to science and technology issues. The exception was the Danish Board of Technology (DBT), which created the consensus conference model specifically for engaging citizens in the assessment of emerging science and technology (*67*).
European achievements in pTA formally entered the U.S. policymaking arena, the birthplace of the TA practice and institutionalization (*68*), through the 2010 launch of the Expert and Citizen Assessment of Science and Technology (ECAST) network, led primarily by the Consortium for Science, Policy and Outcomes (CSPO) at Arizona State University and the Museum of Science, Boston (MOS) (*69*). Building on DBT's World Wide Views Model of cross-national public consultation (*70*), CSPO's research in the social implications of emerging technologies (*71*), MOS's experiments in public engagement in science (*72*), and a decade of innovation and opportunistic applications in biodiversity, planetary defense, climate resilience, autonomous vehicles, and solar geoengineering, and human gene editing, ECAST developed an adaptable and scalable three-phase model of pTA (*73*) consisting of (i) Problem Framing and Design, (ii) Citizen Deliberation, and (iii) Results and Integration. In addition to informing policy and decision-making, ECAST pTA has been applied to improve scientific literacy (*74*), map public values (*75*), and make process innovations (*76*).

Participatory technology assessments can be carried out in all three phases of research and development, focused on different sets of policy and decision-making questions. In the upstream, it involves questions about what R&D to undertake; in the midstream, it involves questions about how R&D should be conducted; and in the downstream, it involves questions about whether and under what conditions the results and outcomes of R&D would be used or implemented (*77*). However, it is most effective when technologies are still nascent, while key questions concerning the design, application, and intended scale of use remain open and can be reshaped by informed, inclusive, and deliberative public input. For these reasons, participatory technology assessments are likely to be quite valuable for early stage fusion energy system designs.

## Speculative design

Speculative design, in contrast, is an orientation that invites users and designers to imagine possible futures as a way to examine the present moment and work creatively toward new paths and trajectories – for these reasons, this approach holds particular promise for fusion energy development.
Sometimes called, "design fiction," the speculative design process works in tandem with imaginative storytelling about future contexts with a focus on technological innovation (*78*). This approach is not bound to any particular version of the future, but is more interested in creating opportunities to see the present moment in a different way so that design can be used to help new futures emerge. In this sense, speculative designers work backwards by asking, what do you want the future to look like (in x-number of years, and in a specific context)? This question can be asked across at least four scenario types, including those that seem *probable*, *plausible*, *possible*, and *preferable* (*79*). The answers to these questions reveal how users feel about the here and now, and create an opportunity to explore what is "working" now, where we are headed if we continue on this path, and if we want to continue in that direction. Speculative design is, then, a tool for analysis and even critique of current designs, processes, or designed circumstances. This approach can be applied in a range of contexts, from art to furniture, architecture, infrastructure, and energy systems design. To the extent that culture and design are intertwined, as described above, this approach can also invite opportunities for cultural critique and analysis. Speculative design can allow users and designers to explore or expose inequities of power that might be installed or maintained by designed products, processes, or built environments and infrastructures. This approach can be particularly relevant to nuclear engineering contexts given the enduring nature of nuclear energy facilities, and even more so, of nuclear waste repositories. In some small trials, using speculative design in nuclear energy contexts has proved useful (*7*), and may help mitigate some challenges, including community resistance or misunderstanding, that often emerge in nuclear siting and design projects.
An important aspect of speculative design is the thoughtful management of "speculation." In this case, designers and users are encouraged to design outside of well-established visions of our technological future (as shown in popular films, for example) (*80*). Instead, speculative design asks for examination and rethinking of current values and beliefs that might have led to our current technological situation and its supporting innovations. In other words, if we designed with different values, beliefs, or priorities in mind, might we have developed different technologies or focused on developing technologies to solve different problems or meet different needs? Further, in speculative design, we can imagine, in these alternative framings, how future technologies might be developed that move communities and society more broadly, off of current trajectories and timelines? This approach is sometimes termed "critical design," emerging in the 1990s out of the Royal College of Art in London. Notably, this approach can sometimes produce strange, implausible or dystopian visions of the future, which has led to some criticism of the approach. Advocates suggest that these strange or unreal imagings can also be quite valuable, encouraging designers to think more critically across the design process and offer a way to critique current values, beliefs, and systems that are not serving a shared common good (*79*, *81*).

## Design justice

Design justice also approaches that design process with critique and critical analysis as guiding aims. Specifically, design justice (*82*, *83*) is a framework that critiques traditional top-down design practices that can often be disconnected from the context of use and thus intentionally or unintentionally exclude users or those most likely to be impacted by the technologies being designed. Design Justice offers strategies for inviting reflexivity and accountability into the design process. It asks designers to consider several questions as they go about the work of designing: Who gets to design and develop new technologies? (And who does not?) Who do we design for and with? How do we scope and frame design problems? What values do we encode or reproduce in the technologies we design? And how do we remember and share stories about how we design, including assumptions made or problems not fully solved during the design process (*84*)? In the past, failure to consider these questions during the design of large technological and complex systems has produced what engineers and designers often refer to as 'unintended consequences'. Yet these unintended consequences are failures of imagination during technology design and development, which can be mitigated through conscious efforts to integrate reflexivity and accountability to end users and impacted populations into design work. According to the design justice framework, education plays a central role in empowering citizens and publics to participate in and lead technology design processes. But returning to critiques of the deficit model described above, education should not be in service of facilitating community acceptance or acquiesce to an idea that does not suit their needs or contexts. Users must be able to opt out, rethink, or otherwise shape design projects from the ground-up. It is through this informed, direct participation that shared values can be discovered and translated into design choices (*83*).

## More than human design

More than human design is an evolving design paradigm that calls for a shift away from traditional anthropocentric approaches to design and technology development to consider non-human actors, including the environment itself as a key 'rightsholder' in the design process. Its origins lie in environmental humanities and geography as well as both Western philosophies (such as posthumanism and new materialisms) and Indigenous forms of knowledge (*85*). More than human design rejects human exceptionalism and treats humans as one entity among many as a way to re-establish more ethical relationships between humans and the environment (*86*). Decentering humans does not negate humans but rather calls attention to the "networks of mutual influence" (*85*) between humans and other species, as well as the systemic effects of design across multiple species and the environment (*87*), all of which share interdependencies (*85*). These effects are otherwise not accounted for in traditional design processes and are treated as ‘externalities’ or, as noted above, characterized as ‘unintended consequences’.  For most energy technologies, externalities may take the form of waste products emanating from the normal operation, including greenhouse gas emissions for fossil fuel-based energy and radioactive wastes for fission. However, energy use and generation create impacts well beyond the site at which the energy is produced. These include, notably, the health and environmental impacts of mining (which all energy technologies require, be it for fuel or for critical minerals needed in the production of the energy system itself). Often, these sites of extraction and waste management do not receive the same level of attention or benefit from the same resources allocated for the energy production site, and as a result, health and environmental impacts of local species (human and non-human) may be experienced for decades.

More than human design researchers have been developing new methods for knowing and relating to non-human actors (*86*). Examples of more than human design approaches include the umwelt sketch where an “umwelt” depiction of the perceptual world of an organism. An umwelt sketch is then a visualization of the overlapping perceptual worlds of different organisms as a way to understand the world from a more-than-human perspective (*86*). Embodied ecological walks, fieldwork in nature, and co-designing from the wild offer another set of approaches to more than human design (*88*). More than human design researchers have also developed tools for representing nonhuman actors (these include non-human personas (*89*) and species passports (*90*), both of which are intended to include the concerns and perspectives of non-human entities in design processes). Other approaches call for noticing different ways of sensing the world and attunement to nonhuman rhythms and senses (*85*), as a way of developing ethical and responsible relationships with non-humans, which may extend to more concerted practices of care toward the environment and other species (*91*).

## Mutual learning and preparation

Across all of these approaches to design, there is a shared acknowledgement that traditional design practices are insufficient in their consideration of community needs, knowledges, and contexts. What is needed, as we note in our introduction, and as these many design practices point to, is an ecology of participation where experts learn alongside community members as they proceed through project design, development, and implementation. Rather than a hierarchy, an ecology of knowledge, experience, and needs can build more robust systems, strategically matched to community values and geographies. Instead of privileging or relying entirely on expert knowledge in energy systems design contexts, a practice of shared or mutual learning and preparation where publics have choice and a voice, may net improved results for communities and designers alike.

A commitment to mutual learning is a recurring feature across the design frameworks presented above. The first reason is tied to persistent knowledge gaps between technical experts and users. Users and experts typically operate within different frameworks of knowledge, resulting in different values and priorities. Values sensitive design addresses this change head-on, as shown above, by inviting user knowledge (in this case, knowledge about what they value) into the design process. However, this process does not close the gap in *technical* knowledge that often exists between experts and users. This problem is most typically addressed by way of education. Of course, as noted above, even the best educational schemes can fall short of their objectives for reasons enumerated in Table 1. Still, education and *translation* work is an important part of facilitating communication in human- and environment- centered design. Technical experts must take seriously the process of translating complex ideas so they are *accessible* and *useful* for purposes of debate, decision-making, and otherwise to support shared communication. Importantly, this work asks experts to think from users' point of view, to find the available means for making information relevant and meaningful. The translation work can serve as a kind of preparation for engagement that may invite in empathy and foster a shift perspective, or new ways of seeing the design problem,  that can be quite valuable in engineering contexts. Mutual learning across differences in knowledge and perspective, and the preparation it takes to facilitate this kind of learning, can make for more robust, context-appropriate design.

A second reason to include mutual learning as a frame for human- and environment- centered design is an extension or modification of the ideas noted above, and tied to the assertion that knowledge is co-produced by users and experts in design contexts. When technical experts translate complex ideas into accessible language, visuals, or other modes of communication, the user must then 'make sense' of those translations. This making-sense process is grounded in users' unique experiences and knowledges. And inevitably, the sense that users make will be something slightly more or new or nuanced than the original technical content (*92*).  If an idea is misunderstood, experts must explain differently, more simply or more expansively. Users take this information and fold it into what they already know. New connections are made, new ideas or new knowledge emerges. In other words, translation and making-sense co-produce new knowledge that can be quite valuable in design. Notably, in this process, neither expert nor user knowledges are privileged or dismissed along traditional lines of bias, but both are needed and valued to successfully understand the design context and move forward.  This practice can facilitate learning about and with users over the long-term can productively reduce this gap and build trust. One recent might might be found in Canada where, over fourteen years, experts in nuclear waste storage of community members living near a proposed storage facility shared their knowledge to build understanding and, ultimately, consensus and consent for moving the project forward (*93*). Which is to say, that preparation and mutual learning is not without its challenges and can take time. Notably, in the context of deep time, fourteen years is blink. One obligation of nuclear engineers is, of course, to think and design for today and across multiple temporal scales.

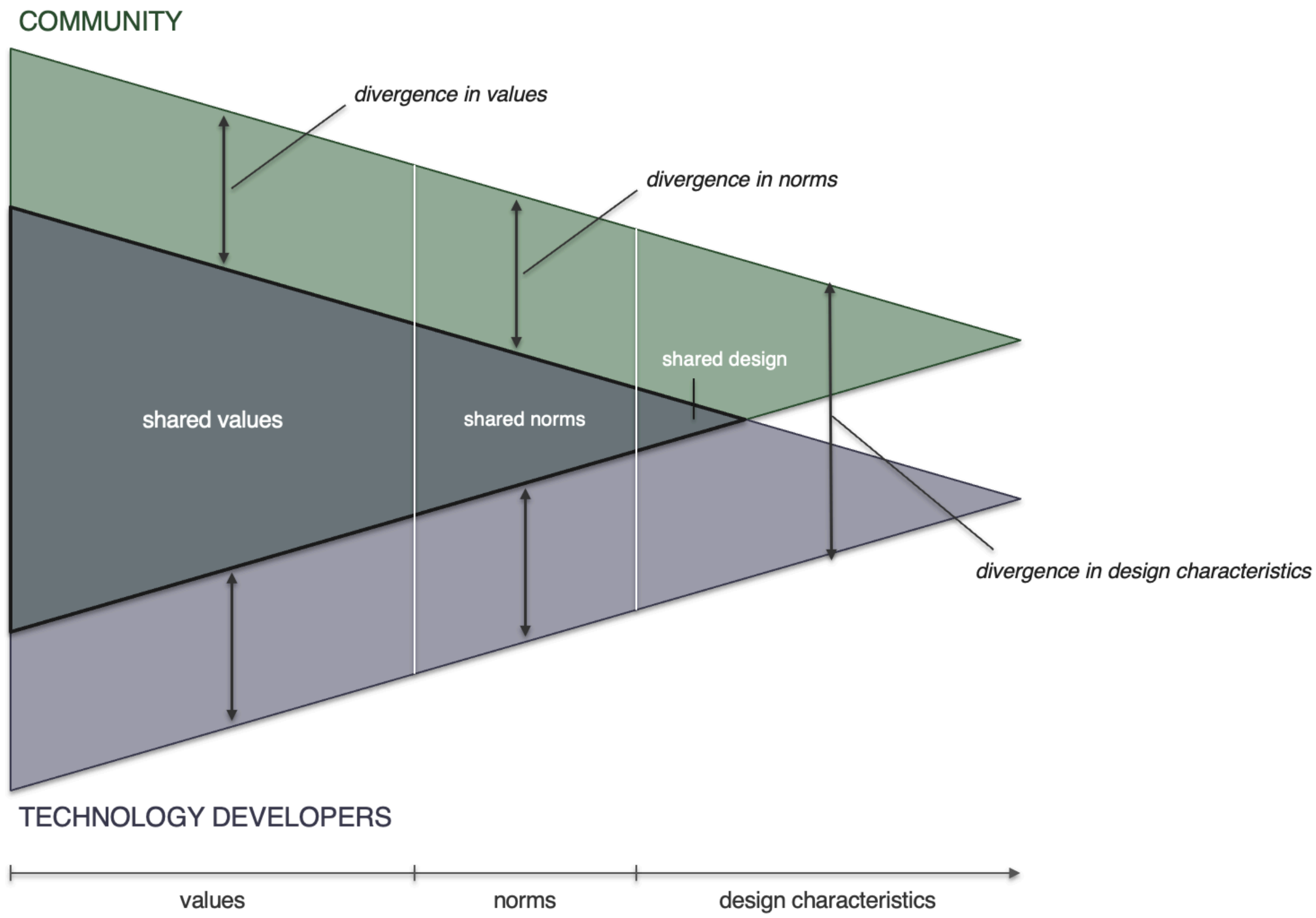

Figure 1. Here is visualized intersections and divergences between communities and technology developers in the contexts of values, norms, and design. Note, specifically, the divergence in design characteristics.

## Interrogating and closing the gap in values, norms and design characteristics

As shown in the design frameworks described above, technology developers and community members may each hold their own sets of values. Some values may be broadly shared by individuals in each of these groups whereas others may be held by individuals. Values can give rise to norms and norms to design characteristics. We place values at the bottom of the pyramid shown in Figure 1 as they form the basis for norms which then become the basis for design characteristics for artifacts or systems. Framed in this way opposition and resistance to a technology can be understood as a mismatch or divergence of values, a mismatch of norms, or a mismatch of design characteristics – as arrived at by designers and community members. A divergence can occur at any of the three levels of this pyramid, with a mismatch in values or in norms leading almost certainly to a mismatch in design. The goal of creating an ecology of participation and a system of mutual learning is for publics and communities, through multiple encounters, to discover and articulate this mismatch and where possible, seek to close it.

We view this work as necessarily being an ongoing exercise and unlikely to be completed or achieved through episodic, one-off engagements. It is worth noting of course that ultimately, it may not always be possible to close the divergence and some technologies, imagined in any form are likely to be viewed as being fundamentally at odds with societal norms. Indeed international laws and conventions have identified more than one instance of such technologies.

In the sections that follow, we describe our own efforts, through a series of case studies to explore, articulate, and where possible, close these divergences in values, norms, and designs.

# Case studies of human and environment-centered approaches to fusion energy

In the six case studies that follow, which are taken from our work, we describe how both human and environmental perspectives can inform the design of fusion energy systems.

## pTA focus groups

A series of open-framing focus groups was launched in 2023 to engage community members early on during the design phase of fusion energy systems. These open-framing focus groups were adapted from the larger Participatory Technology Assessment (pTA) methodology developed and fielded by the Expert and Citizen Assessment of Science and Technology (ECAST) network (*73*). While the ECAST pTA has three phases – framing, deliberation, and results integration – only part of the framing portion was adapted for this effort. Traditionally, the pTA Problem Framing portion includes: literature review, open framing focus groups, and the stakeholder design workshop. Great emphasis is placed on empowering citizens to speak through their lived experiences during the open-framing focus groups. This practice allows researchers to better understand their general hopes and concerns regarding a technology within its broader societal context, thereby opening up, rather than closing down, sociotechnical options (*94*). Research involving open-frame focus groups has revealed it as a valuable approach to bridge the expert-lay divide (*95*). The information gained through this initial phase is then used to inform an expert and stakeholder workshop for deliberation design.

As an initial application of the pTA process, a series of open-framing focus groups was launched and framed around energy transitions in the state of Wisconsin – a state that is becoming a hub for fusion energy development. The purpose of this effort was twofold: 1) to obtain participants' hopes and concerns for fusion energy systems and 2) to provide an avenue to build trust and transparency through the early phases of the design process for a large sociotechnical system.

Two variations of the open-framing focus groups were explored for a total of four sessions (two virtual and two in-person). The focus groups were 6-8 hours in length and provided compensation for participation. The general layout of each variation asked participants to consider renewable energy, fossil fuel, and nuclear fission as energy sources to consider during an energy transition. Once the participants had given feedback on each form of energy, the facilitator introduced fusion energy as a new form of emerging energy technology to be considered, provided a brief presentation on fusion energy, and offered a brief explainer that gave background on fusion energy and a comparison of fuel consumption for each energy source to power a city. The first variation framed the energy transition discussion in terms of environmental impacts (land, water, and air), and the second variation was framed in the general concept of considerations for energy choices.

Prior to the open-framing focus groups, each participant filled out a consent form and an extensive survey providing demographic details as well as their opinions on fossil fuel, renewables, and nuclear energy sources via 5 point Likert items (very negative, negative, neutral, positive, very positive) and their general level of concern with both energy and environmental issues via 4 point Likert items (not concerned at all, slightly concerned, moderately concerned, highly concerned) . These questions were included to ensure the overall applicant selection and their individual focus group table assignments covered a wide range of identities and perspectives. In both cases, participants were also asked in their initial applications to share "a short story about an influential experience (good or bad) nature has had on your life (5-7 sentences)." This was included primarily to ensure that applicants were willing to provide a reasonably thoughtful response to an open-ended question operating on the premise that those who would not put effort into answering an open-ended question on the application would be less likely to provide thoughtful responses during the workshop if they were selected to participate. Selected participants were provided a workbook to record their responses and assigned to smaller breakout groups (3-5 people) with a facilitator and notetaker.

Post survey results indicated near-unanimous participant agreement, that they had been treated respectfully, been given opportunities to voice opinions and concerns, and that the organizers had effectively moderated discussions and created an open atmosphere for sharing perspectives. Several participants praised the small-group discussion format. One concern, expressed by the participants of the in-person open-framing focus group, was the insufficient time to process educational materials before engaging in discussion.  Several participants expressed an interest in learning more about fusion energy systems. Particular areas of interest included learning about realistic timeline for fusion energy technology development, potential safety concerns and risks, costs of fusion energy in the future, and alternative applications of fusion research. There was also a strong interest in learning about who is making key decisions about how fusion energy systems are being developed. Participants expressed a strong enthusiasm for staying connected with the research team beyond the open-framing focus groups (*96*). This, along with a similar interest expressed by the participants of the participatory design workshops described next, was our main motivation for creating the Global Fusion Forum described below. Given the experiences of the participants, we consider these open-framing focus groups to be a success and a beneficial precursor to future participatory design efforts as they help generate valuable information about public framings of the sociotechnical context for fusion, comprehension of educational materials and elicit additional topics of interest, as well as uncover public hopes and concerns that can inform subsequent participatory design engagements.

## Participatory design workshops

In the Fall of 2023, Verma and Snyder taught a course on community-engaged design of energy technologies (*97*). The entire course was framed as a living lab (*98*). The students enrolled in the course (first-year undergraduates) learned the fundamentals of nuclear science and engineering, alongside technical communication, qualitative research methods, and participatory design – all toward developing a sociotechnical engineering competence as shown in Figure 2. Throughout the semester, the students developed their participatory design skills and prepared to engage with community members initially by interviewing each other, as well as friends and family members. They also created virtual tours of fusion facilities (more on this below) intended for a broad audience. These initial efforts gave the students a framework for respectful engagement with community members.

The course offered two structured opportunities for community engagement. The first opportunity was a virtual workshop which focused on questions of process. In this workshop, along with students and community members, we explored perspectives on participatory design processes, including interest and willingness to participate in such processes, potential barriers that might limit participation, as well as who would be considered a trusted representative for those who were unable to participate. The second, and most significant, engagement took the form of an in-person participatory design workshop in which students and community members worked in teams to design a hypothetical fusion energy facility for Southeast Michigan. The students had formed teams by this point in the course and were joined by community members. Each student on a team had a designated role, including that of a facilitator, storyteller, archivist, and participant-observer. These roles were central to facilitating and documenting the team's design process and outcomes.

The day began with a session that framed the workshop goals and set norms for the participants to engage respectfully and collaboratively with each other. Participants (students and community members) who were already in teams also introduced themselves and shared their special skills and expertise that they each believed would be important for their work together. Hearkening back to the role of values in the design process, each team member also shared values that were important to them and which they believed should inform their design work. Teams generated a shortlist of shared values through a voting process. As the workshop progressed, teams were introduced to fusion energy and invited to share and discuss their questions, hopes, and concerns. From here, teams shifted to an active design phase, which started with idea generation, followed by mapping potential design ideas onto a simplified map of a fusion energy facility. Teams then self-generated and selected decision-making criteria for downselecting from a large set of ideas. Through this down-selection process, each team developed a design concept which they first described textually. Teams then used their textual descriptions of design concepts to generate images of their proposed facility designs. Several of these design concepts are shown in Figure 3. Students in the course had previous experience with AI image generators and so understood the level of detail needed for textual prompts in order to generate accurate images. Each team then created a brief presentation documenting their design process and outcomes (an effort led by each team's storyteller) and shared their design concepts with the whole group. The day ended with teams sharing brief takeaways on the workshop as a whole through a two-word "check-out."

Following the completion of the course, four undergraduate students continued working with Verma and Snyder to digitize and analyze the workshop data which included hundreds of design ideas written down on post it notes (and placed on a map of the hypothetical fusion energy facility), participant workbooks (which included individual reflections), team presentations, and responses to Menti questions periodically posed to the group during the workshop. Taken together, when analyzed, these different forms of design data helped us understand and contextualize the team's priorities and design choices. We noted a number of unique characteristics across the teams' fusion facility designs and also observed shared patterns.

Notably, teams exhibited an interest in capturing the history and legacy of the community in the design of the facility, an interest in respecting the facility's natural surroundings and integrating the facility into nature, transparency and access to the facility by the host community, care for the workers, and a priority on the health and safety of the community. Examples of facility designs included the following three. A team from Detroit, which called itself "From Cars to Stars," had an interest in revitalizing abandoned industrial infrastructure; they explored the possibility of repurposing buildings in Detroit to create a new energy industry in the region. Another team, which was particularly interested in operationalizing transparency in the facility design, designed a (quite literally) transparent facility that enabled community members a direct line of sight into the goings on at the fusion energy facility. The facility was also designed to admit visitors. In a third, particularly intriguing design concept, a team chose to playfully integrate the facility into its natural surroundings by using ziplines as a means of conveyance from one facility building to another. These are but a subset of the ideas that were developed by our teams of students and community members (*7*).

Overall, this process demonstrated for us the tremendous potential to learn about community preferences, as well as hopes and concerns, through a participatory design process. We are also of the view that such participatory design engagements can inform real-world fusion energy facility design efforts and support the development of contextually appropriate fusion energy technologies. Another notable insight from our own participatory design workshop is that host communities are likely to have unique preferences for the design of a local energy facility. Given this, fusion developers should strike a balance between designing systems that are standardizable while also being adaptable to local context, from an aesthetic and functional standpoint.

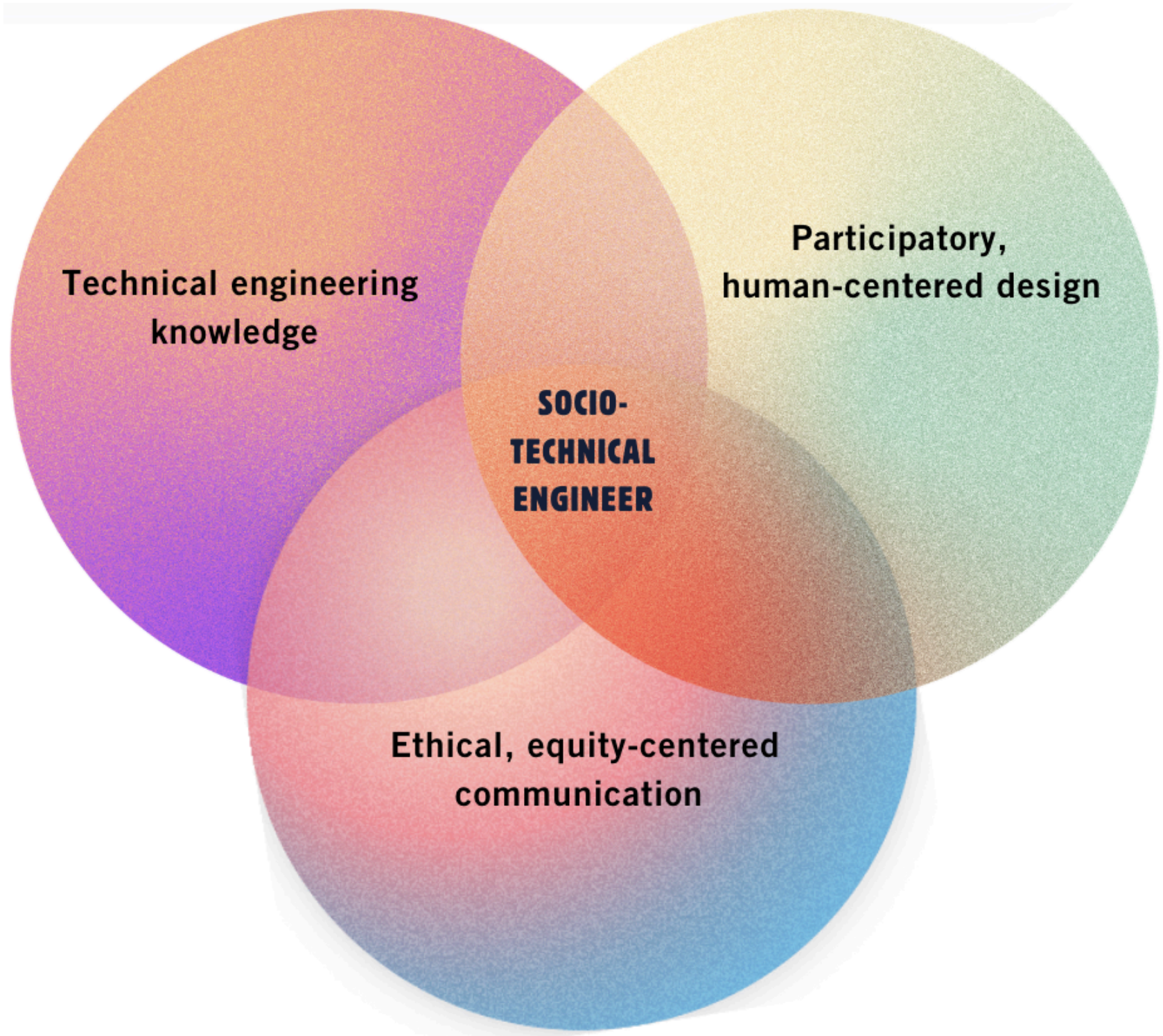

Figure 2: A sociotechnical engineer has excellent technical engineering knowledge, centers ethics and communication in their practice, and uses participatory, human-centered design methods.

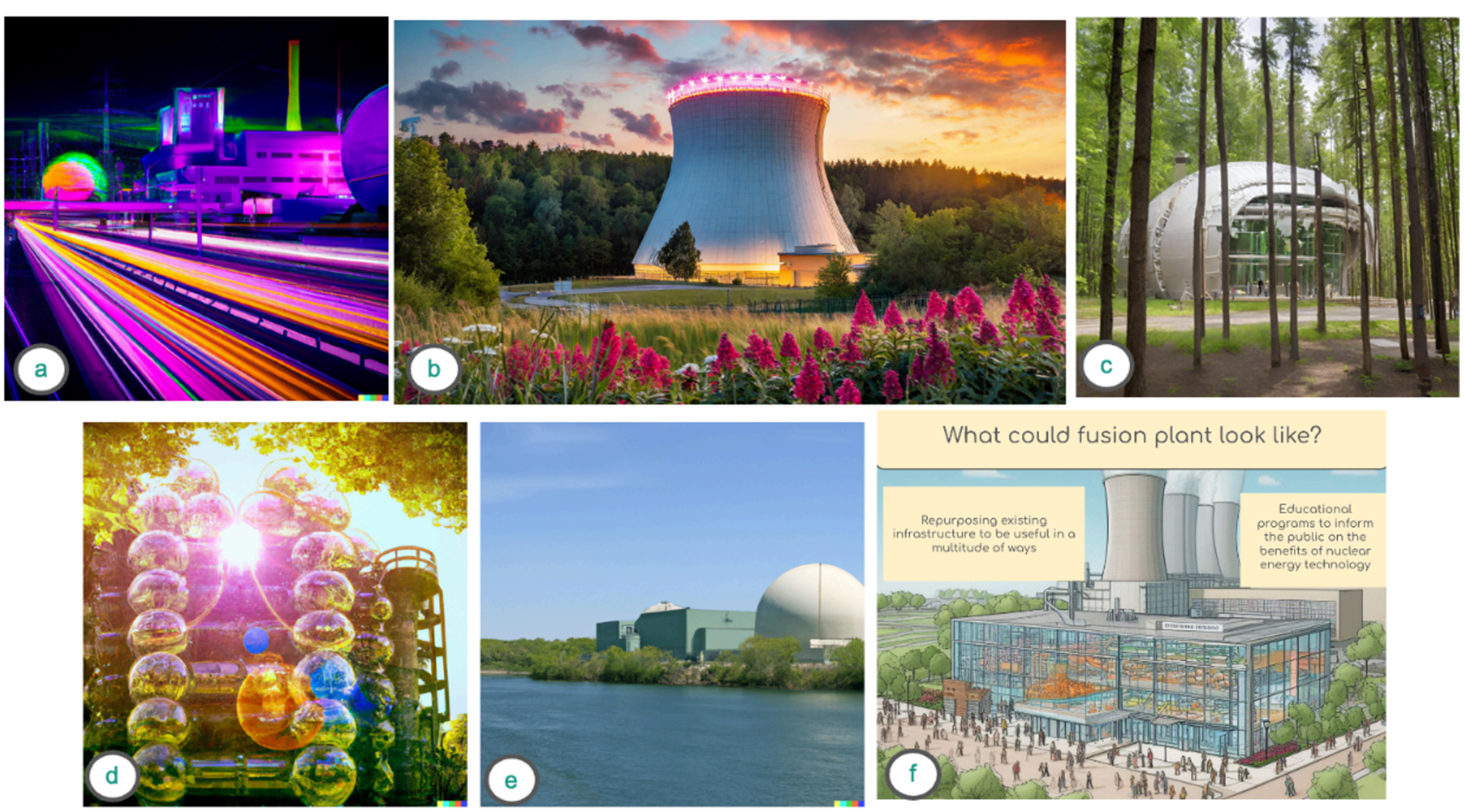

Figure 3: (a) a nuclear fusion facility with colorful lights, sidewalks, and a bullet train rail; (b) A modern designed nuclear reactor with LED lights on it at sunset with nature and flowers surrounding it. Forest engulfs the background; (c)a futuristic fusion energy facility surrounded by trees; (d) A nuclear fusion facility with lots of light and bubbles and surrounding greenery; (e) A nuclear fusion facility with giant windows on a river; (f) A transparent fusion nuclear energy plant surrounded by plant workers (the students superimposed the text onto the image). Adapted from (*7*, *32*).

## Immersive experiences of fusion energy technologies

Although many companies are working toward the rapid commercialization of fusion energy, no fusion energy facilities currently exist. And while there are a number of fusion experimental devices around the world, they can be difficult to visit. It was our goal to create an immersive experience of a fusion energy facility for our students as well as members of the public who might be interested in the technology. To that end, we have created VR models of a tokamak-based facility modeled largely after ITER (ITER being the only design for which we could reliably find external and internal design details). More precisely, we have created two sets of VR models. The first model offers a cross-sectional view of a tokamak that users can explore and label as a way to learn about the main components of a tokamak-based fusion energy system. The second model is designed as a first-person player video game in which the user can walk (or fly) around the fusion energy facility to better understand the scale of the fusion device and explore how it fits within the broader energy facility, and how that facility itself might fit into the environment and landscape of the broader community.

In replicating the tokamak-based facility in Virtual Reality for accurate and immersive visualization, our development team used various 3D software and consulted experts and official ITER resources. First, a 3D model of ITER was sourced from the ITER Communication team and Centre for Energy Research, initially created by Tamás Szabolics and Dénes Oravecz and published for educational use. The ITER model that was acquired was highly complex, featuring all of the external and internal components and modeled in such a way to allow students to view a cross-section of the reactor in order to better understand the reactor assembly. Therefore, to be compatible when streamed to a VR headset, our student artist optimized the model in Rhinoceros which involved greatly simplifying its digital geometrical surface. Simultaneously, she modeled the surrounding facility using 360 photographs and building maps published by ITER, and reviewed the accuracy of its scale and layout by individual testimonials. Once combined in Unreal Engine 4, students were able to experience ITER in complete context, and explore the full facility. They were guided by a teleport menu with various points of interest. Additionally, within Unreal Engine 4, a companion Virtual Reality app was developed to challenge students to identify and label the different system components on their own. Students navigate the Virtual Reality model using a fleet of University-provided Meta Quest Pro VR headsets. These headsets operate tethered to high-powered workstations, leveraging the RTX 4090 GPU to render the VR scene in precise detail and allow for expansive exploration of the facility. Figures 4, 5, 6, 7, and 8 show cross-sectional facility level models of ITER throughout the process of development of the final VR models.

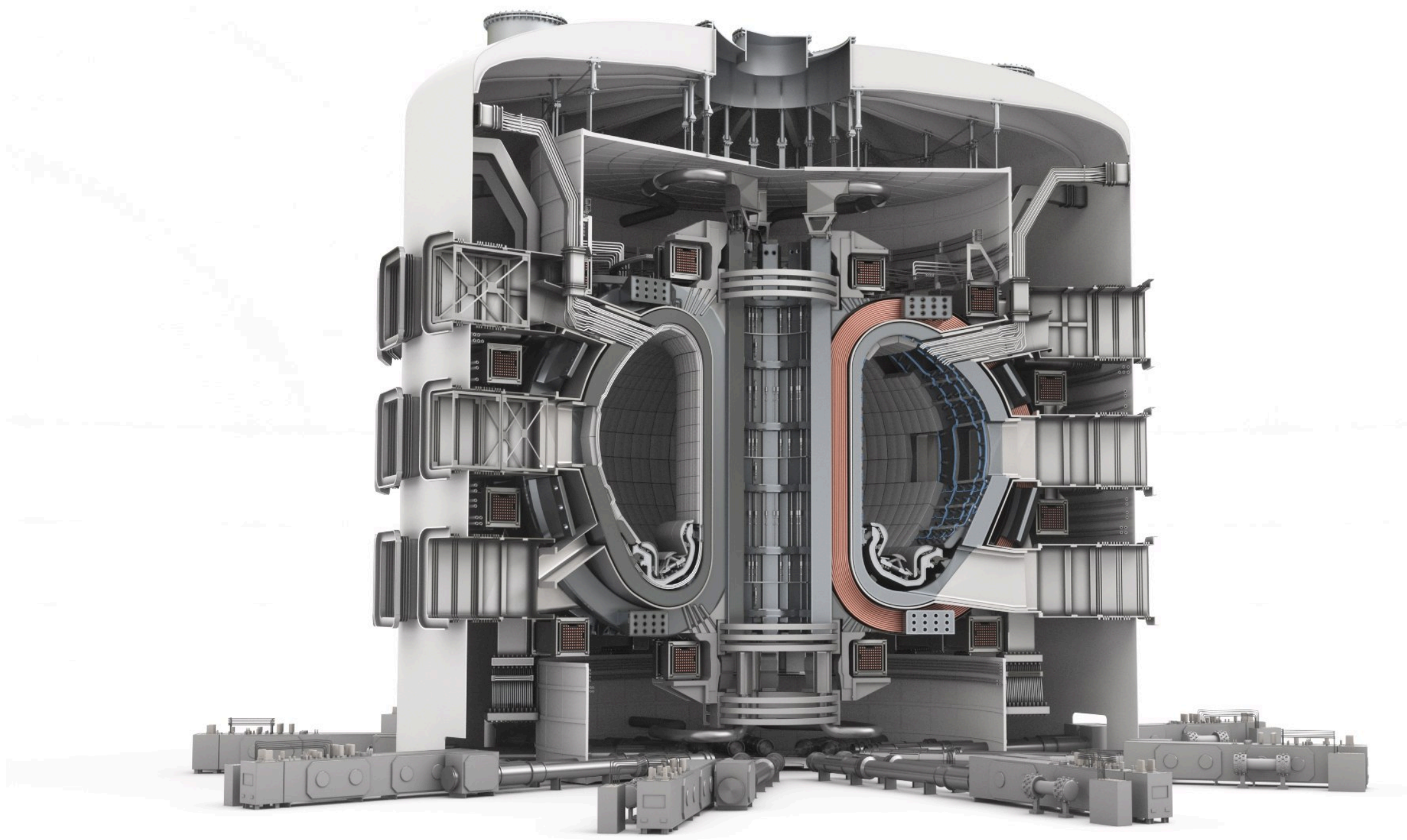

Figure 4: Complex ITER model prior to optimization for Virtual Reality streaming.

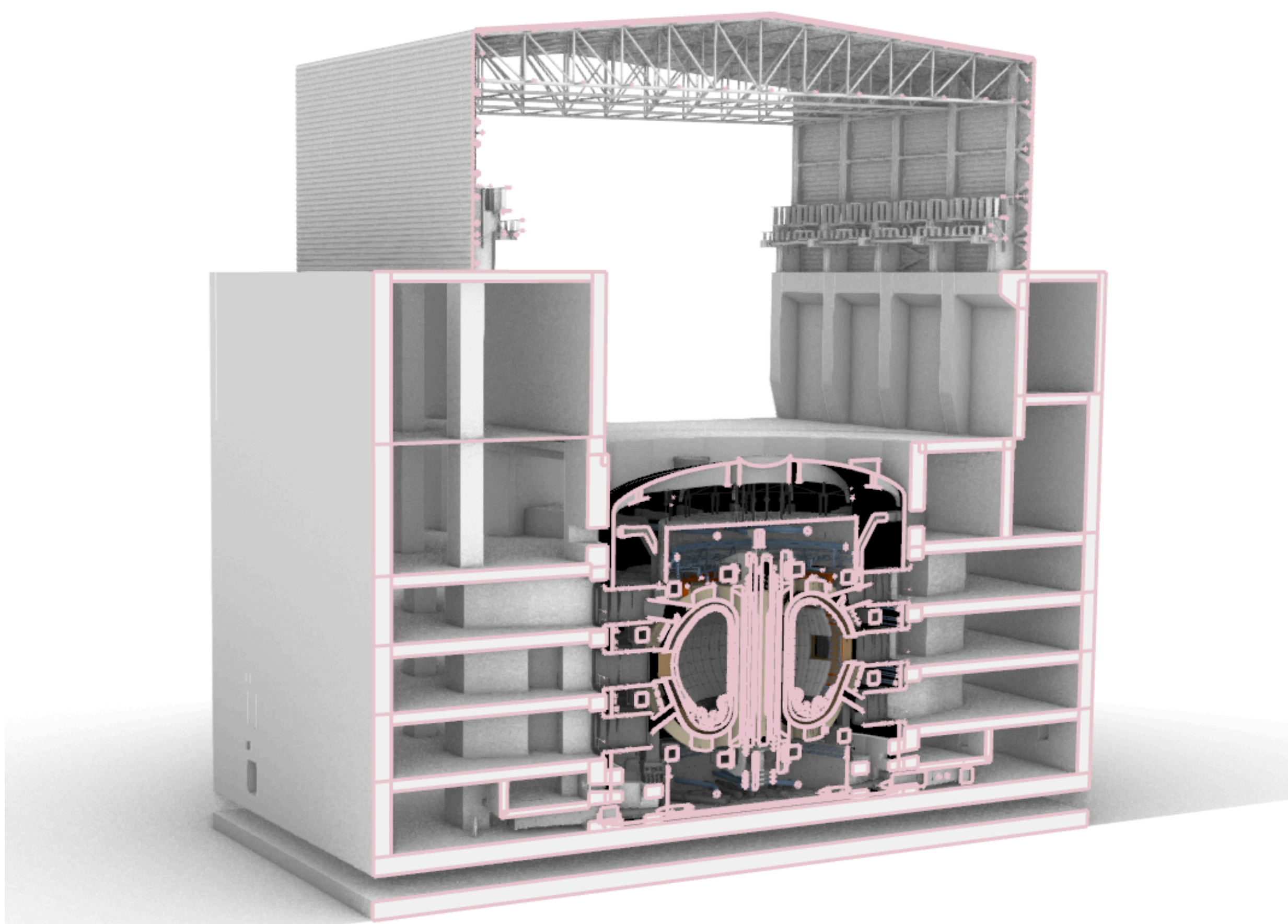

Figure 5: The Cross-Sectional view of ITER in Rhinoceros with in-progress assembly hall.

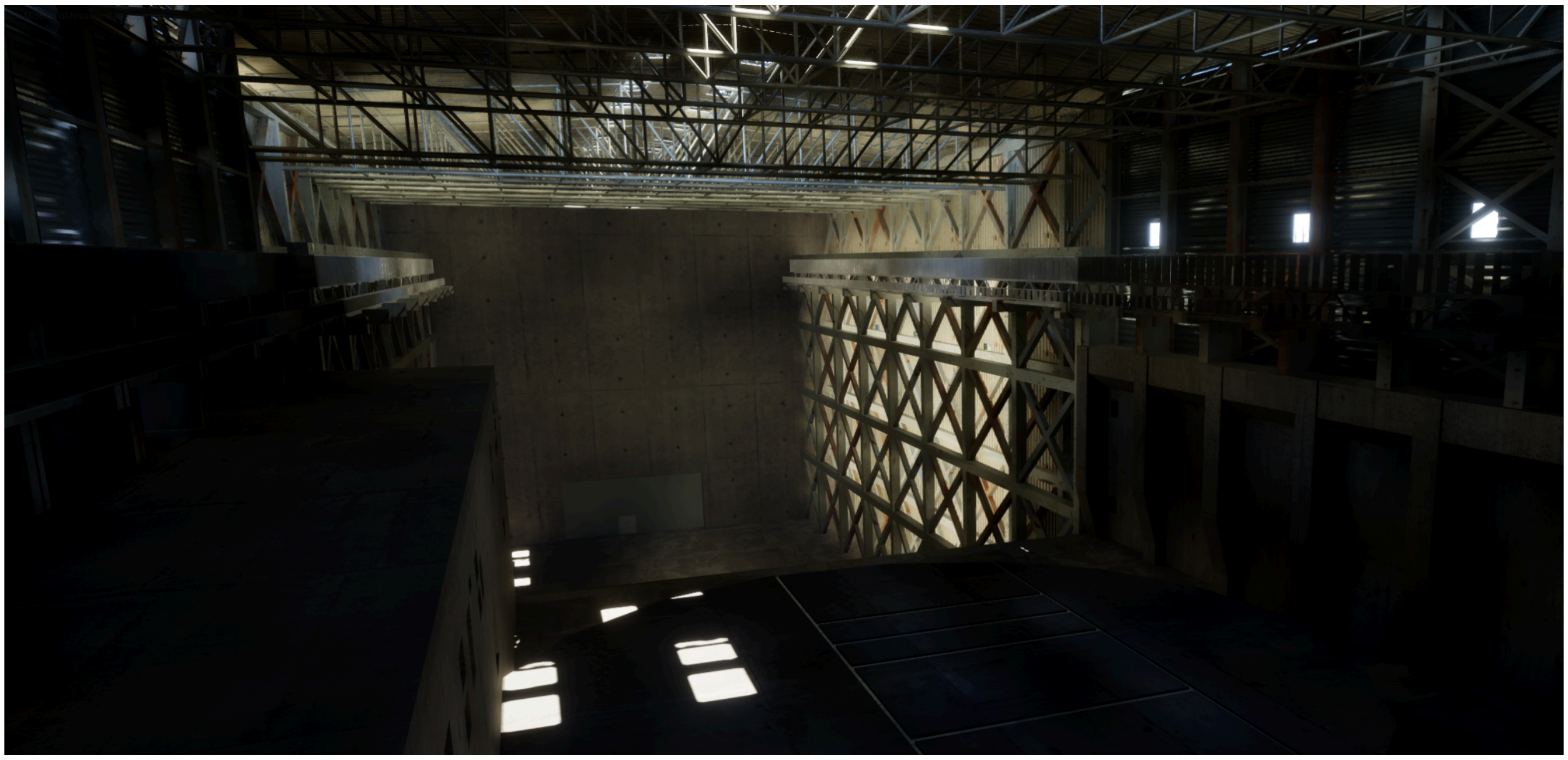

Figure 6: Assembly Hall constructed in Rhinoceros displayed in Unreal Engine.

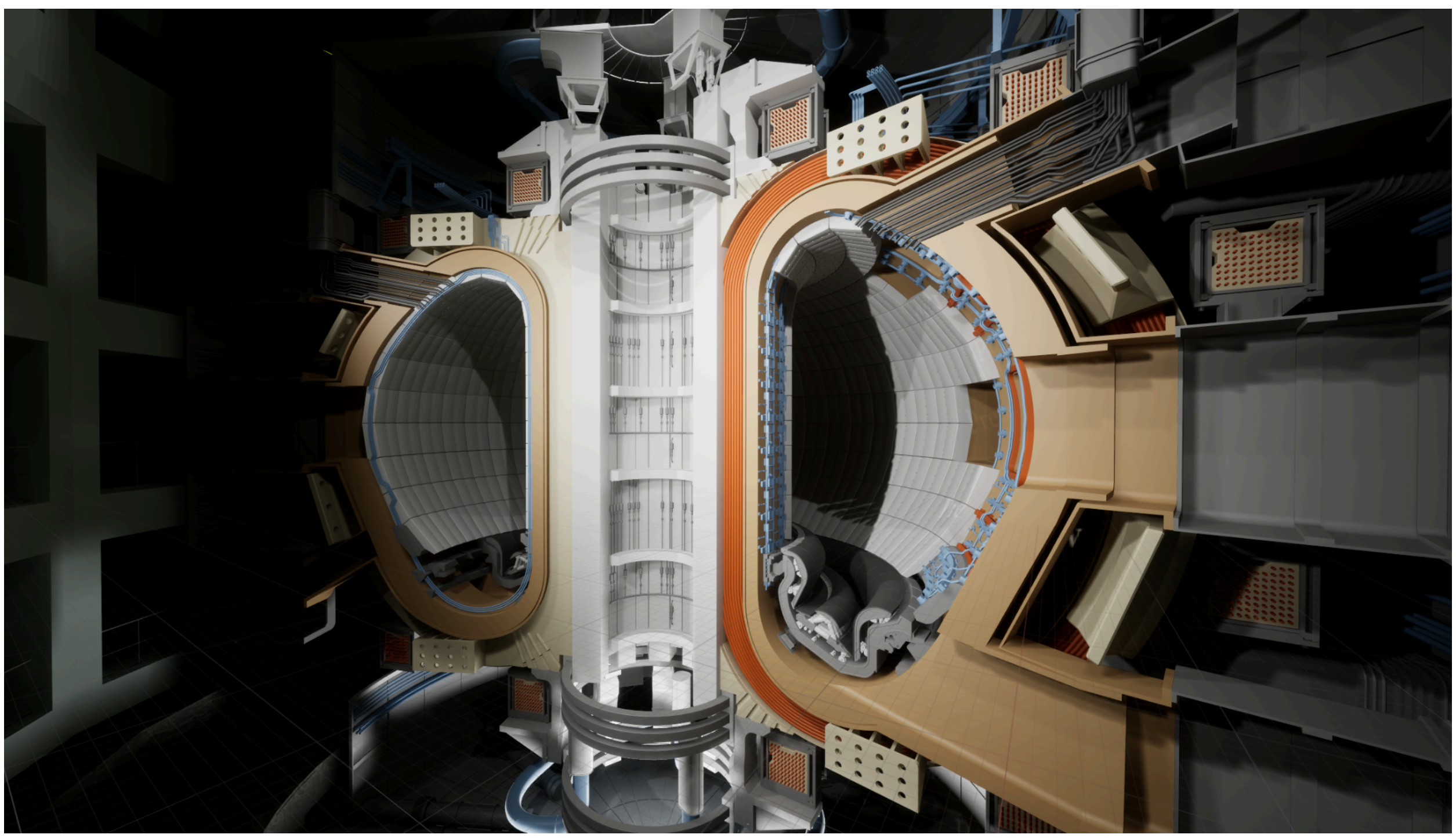

Figure 7: The optimized ITER system from the perspective of a student in VR.

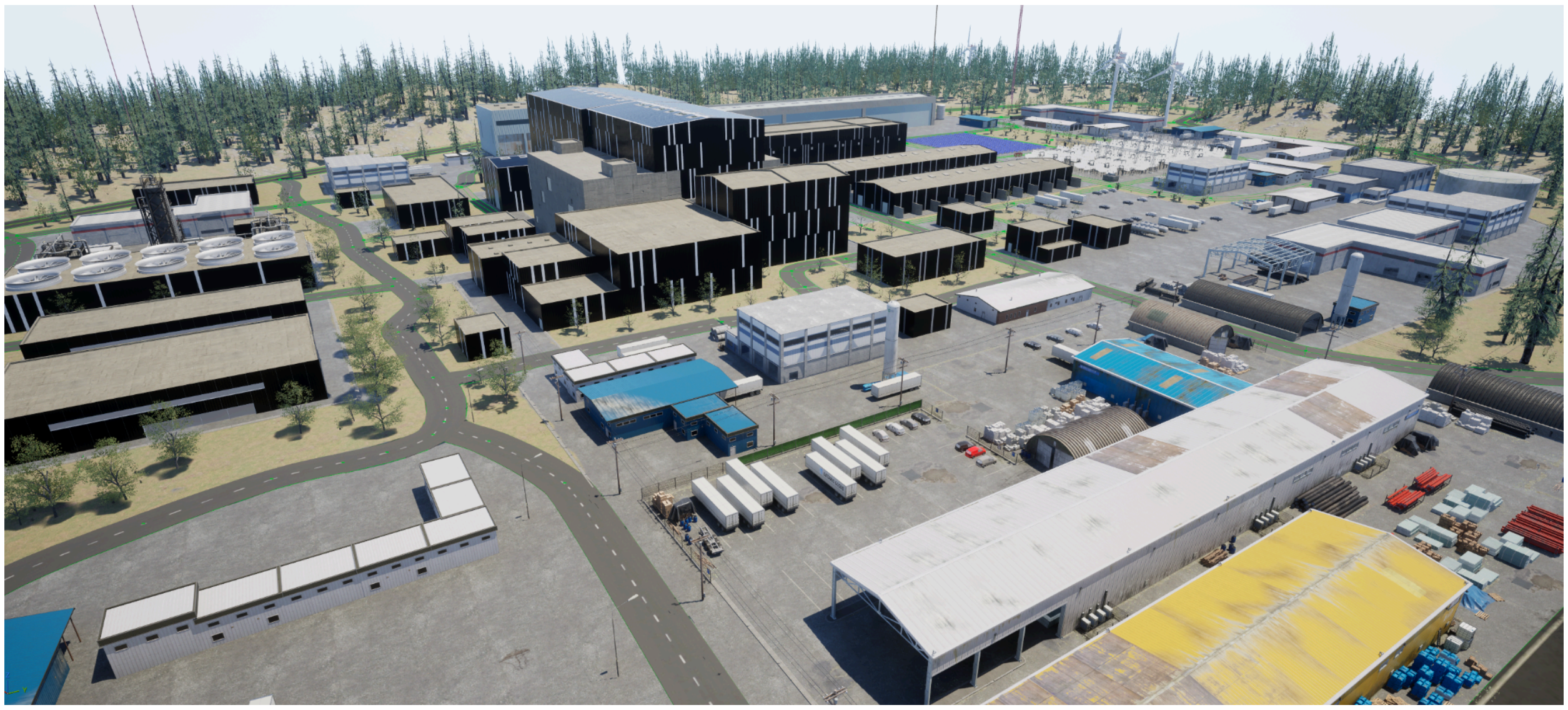

Figure 8: Aerial view of Campus created in Rhinoceros and experienced in VR based upon the campus map published by ITER.

As part of one of their assignments, the students in the undergraduate course were asked to create a tour of the fusion facility using the VR model. To do this, students first explored the model themselves and then developed a script for the tour, keeping in mind their intended audience and the questions and interests of such an audience, including what parts of the fusion energy facility the tour group might be interested in visiting. Students then screen-recorded a visit to the facility using their script. In an unexpected development, many students shared these visit videos with their friends and family members, thus creating a second unintended (by us) but delightful opportunity to share their knowledge with their own communities (*97*, *99*). Following this assignment, students were invited to consider a series of reflection prompts as shown in the table below, along with representative responses.

Table 2: Student reflection prompts and sample comments following interactions with VR reactor models

| |
|---|
| 1.What is your initial reaction to the VR model you have just explored? (no wrong answers. Please respond in a couple of sentences<br><br>*"This reactor was much more the size and shape I was expecting. It was amazing to see how they fill every bit of space inside with some piece of equipment. When seeing these things normally on the computer, I tend to imagine a lot of empty space, or just disbelief as to how this exactly happens, so it was very interesting to start to see how it all fits together. I also really like the different approach that this one took to highlighting the features of the reactor by taking a slice through the middle as opposed to taking apart layers. I thought was a much better approach for this kind of reactor."* |
| 2.What are your initial reactions about the scale, scope, and complexity of these systems?<br><br>*"Going into this class I already had a base level understanding of the complexity of a fission reactor but upon seeing the complexity of the fusion reactor I was dumbfounded. It really opened my eyes to just how advanced technology can be and was very fun to explore and a little intimidating to try and understand how it works."* |
| 3.Are there any aspects of the systems that are particularly intriguing to you? (if so, which ones, and why?)<br><br>*"I thought the blanket and the vacuum vessel were really intriguing. The tunnel they made around the core of the reactor was cool to walk through in VR because it allowed me to see what the inside of the reactor looks like."* |
| 4. Now imagine that this system is about to be built in your hometown. As a resident of a town that might host this facility, what questions, concerns, or hopes might you have about the development of this system?<br><br>*"If a fusion reactor was built in my hometown I would definitely want an explanation of how it worked and how safe it was as I doubt I would trust it seeing how experimentally new it is. I too would have high hopes for cheap electricity and hope it would not nearly be the eyesore the fission plant is."* |

In general students expressed feelings of either overwhelm or fascination regarding their initial encounter with the ITER system in VR. Those feelings of overwhelm quickly turned to curiosity about issues like durability of materials, complexity of design, and scale in comparison to existing fission energy systems. Primary areas of concern were tied to uncertainty about establishing safe operating conditions, questions about the longevity of the system, actual power production, costs, and safe disposal of radioactive waste.

These VR models have now been used by undergraduate students for over three years. They have also been used by three cohorts of summer school students, a group of visiting students from Tohoku University, and participants at the inaugural Global Forum Symposium (held by the University of Michigan in October 2025).

## An online platform for learning about and informing fusion energy technology design

A number of community members who participated in our design workshop as well as the participatory technology assessment framing groups expressed an interest in continuing to learn about fusion energy and staying engaged with us. Throughout the workshop, we also heard strong expressions of interest from the participants in learning about fusion energy and helping shape key technology decisions – particularly those having socioeconomic and environmental impacts. At the same time, we were aware of growing survey-based research efforts in Europe and well as the US, whose main findings indicated a positive sentiment towards fusion energy while noting that that positive sentiment was rooted in an incomplete or low level of understanding about fusion (*57*, *100*). Given the interest in learning about and informing fusion energy decisions from our participants, coupled with broader evidence of low fusion literacy, we believed it was important both for interested publics to be able to have a say in the development of a potentially significant energy technology (from an energy security, access, and geopolitical perspective) and for this say to be rooted in an understanding of the technology itself. These dual motivations led us to design an online platform that interested individuals can use to both learn about fusion energy while also expressing their perspectives on how fusion energy technologies ought to be developed. The platform, which we launched in May 2024, at the same time as the first international Fusion Energy Week (in collaboration with the US Fusion Energy Team), is called Global Fusion Forum.

The development of the platform combined expertise in plasma physics (Diem), nuclear engineering (Verma), technical communication (Snyder), user experience and interface design (Morales Coto), scientific illustration (Kova), and web-application development (Hoelzle and Daley). As the team developing this platform, we first started by identifying potential users of this platform. We imagined prospective users as including community members who had participated in our participatory design workshops and participatory technology assessment focus groups, interested publics in the US and internationally who wanted to learn about fusion energy, communities in the US who host energy facilities, communities in the US who might be interested in hosting fusion energy facilities, environmental justice communities, and K-12 students. We then developed a list of questions that each set of users was likely to have about fusion energy. This list of questions was generated based on our knowledge of each set of users through our direct interaction with them in prior engagements. We then merged and reframed some questions until we arrived at a shared set of questions about fusion energy, which would be addressed on the Global Fusion Forum platform. Kova developed scientific illustrations for each of the key thematic areas represented in the Global Fusion Forum. The illustrations are intended to live in the sweet spot between scientific accuracy and poetic interpretation, creating a sense of energy and warmth. Some examples of these illustrations are shown in Figures 9.

To develop the written content, Snyder, Verma, and Diem worked together to shape technical content into short, easy-to-read chunks. Diem and Verma initially provided longer narratives explaining key aspects of fusion technology design and innovation, then Snyder condensed the content to fit the design space and translated technical ideas into more common or familiar ideas and phrases as appropriate. Verma and Diem reviewed the shortened content for accuracy and made suggestions for changes. This revision process went back and forth for several iterations, including consultation with designers Hoelzle and Daley, before content was finally approved for inclusion in the application.

Users visiting the Global Fusion Forum website for the first time are asked a few simple questions about themselves (as shown in Figure 10) without revealing any personally identifiable information (unless they choose to share an email address as noted below). These include questions about their role (are they a student, educator, community member, elected official, industry employee, member of a non-governmental organization, or other). They can also share their home country and state, as well as four values that most matter to them (here users can select values from a predetermined list of twelve values. This list was populated based on findings from the participatory design workshop described above). Users who are interested in participating in future design workshops or participatory technology assessment efforts can share their emails so that they can be contacted at a later date when a design workshop or PTA event is held near them.

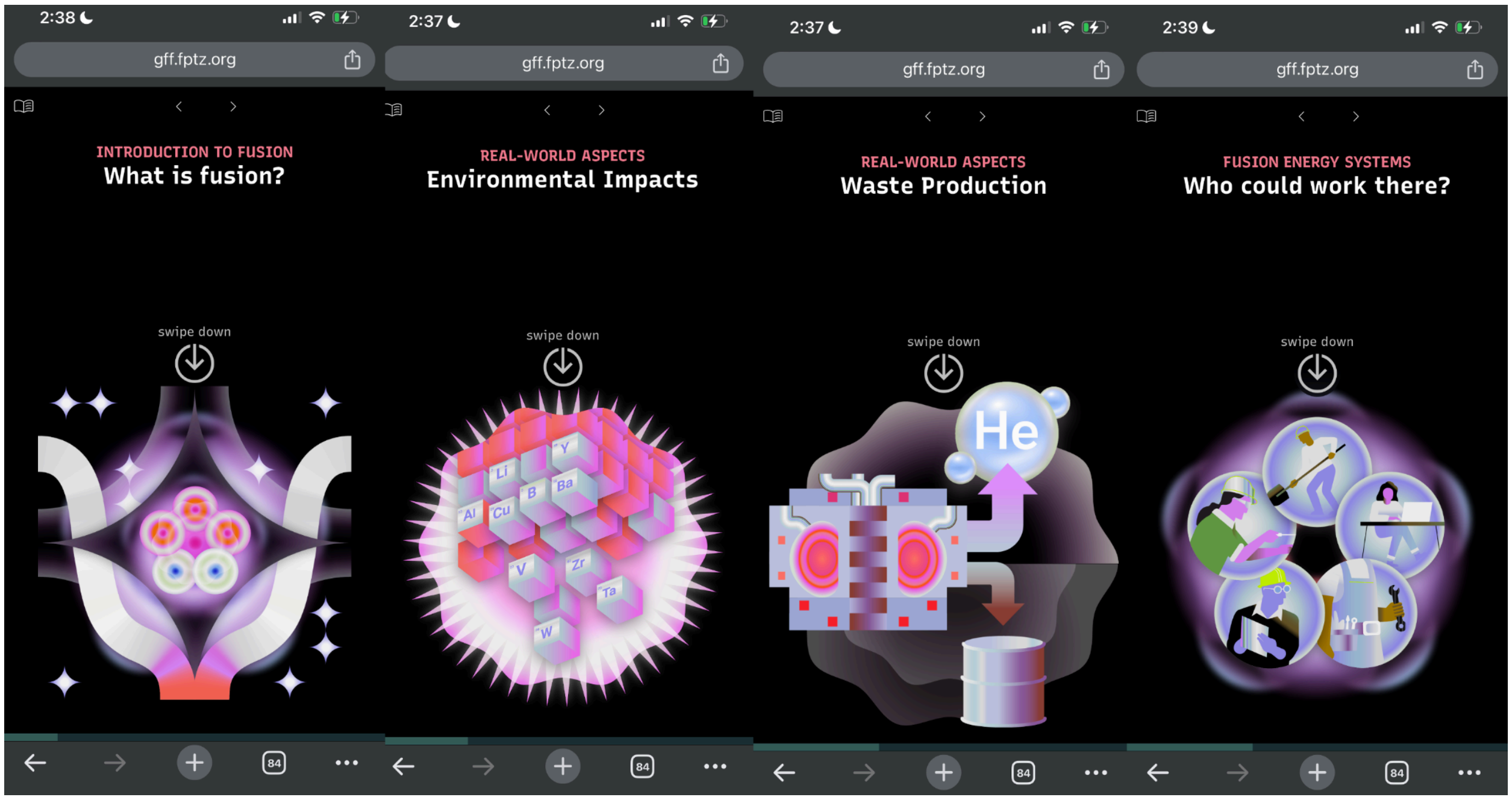

Figure 9. Examples of original artwork developed for the Global Fusion Forum

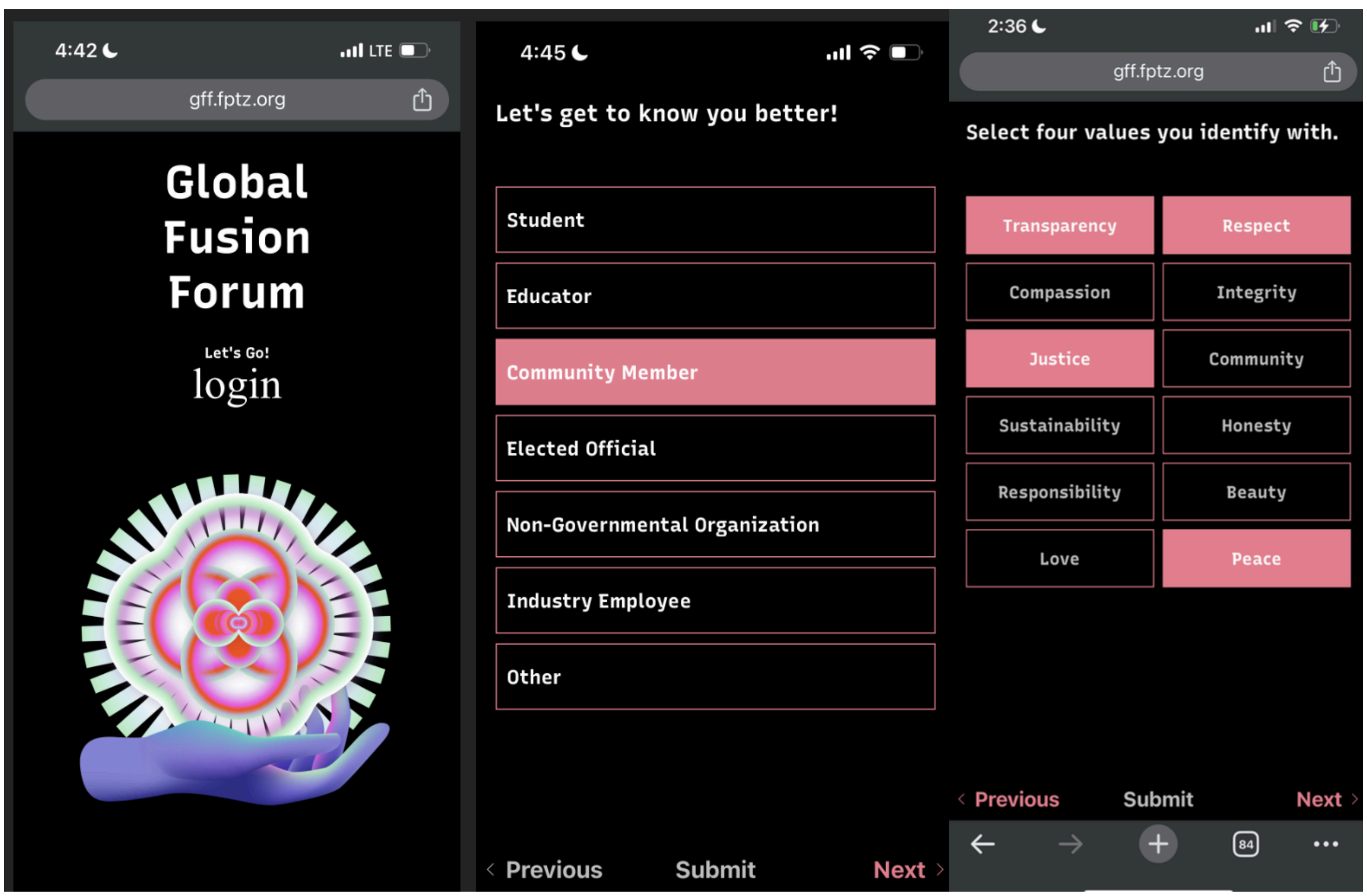

Figure 10: When first accessing the Global Fusion Forum, users are asked a few simple questions about themselves without being asked for any identifying information

Information on fusion energy is organized under three broad themes: (1) An introduction to fusion (what is fusion, the conditions needed for fusion, and common myths); (2) Fusion Energy Systems (What is a fusion energy system, how are fusion energy systems built, Who could work at a Fusion Energy Facility? (Figure 11)); and (3) Real world aspects (Anticipated timeline for fusion energy development, regulation of fusion, waste production, and environmental impacts (Figure 12)). Selected examples from these sections are shown in Figures 11 and 12. As users learn about each topic, they are periodically asked questions. These questions are not meant to test their knowledge but rather are a way for the users to share their own perspectives on what they are learning. For example, in one question, users are asked about which aspects of fusion energy development they believe require the most research. In another question, users are asked whether they would be interested in working at a fusion energy facility. Since its launch in 2024, Global Fusion Forum has been used in over 40 countries.

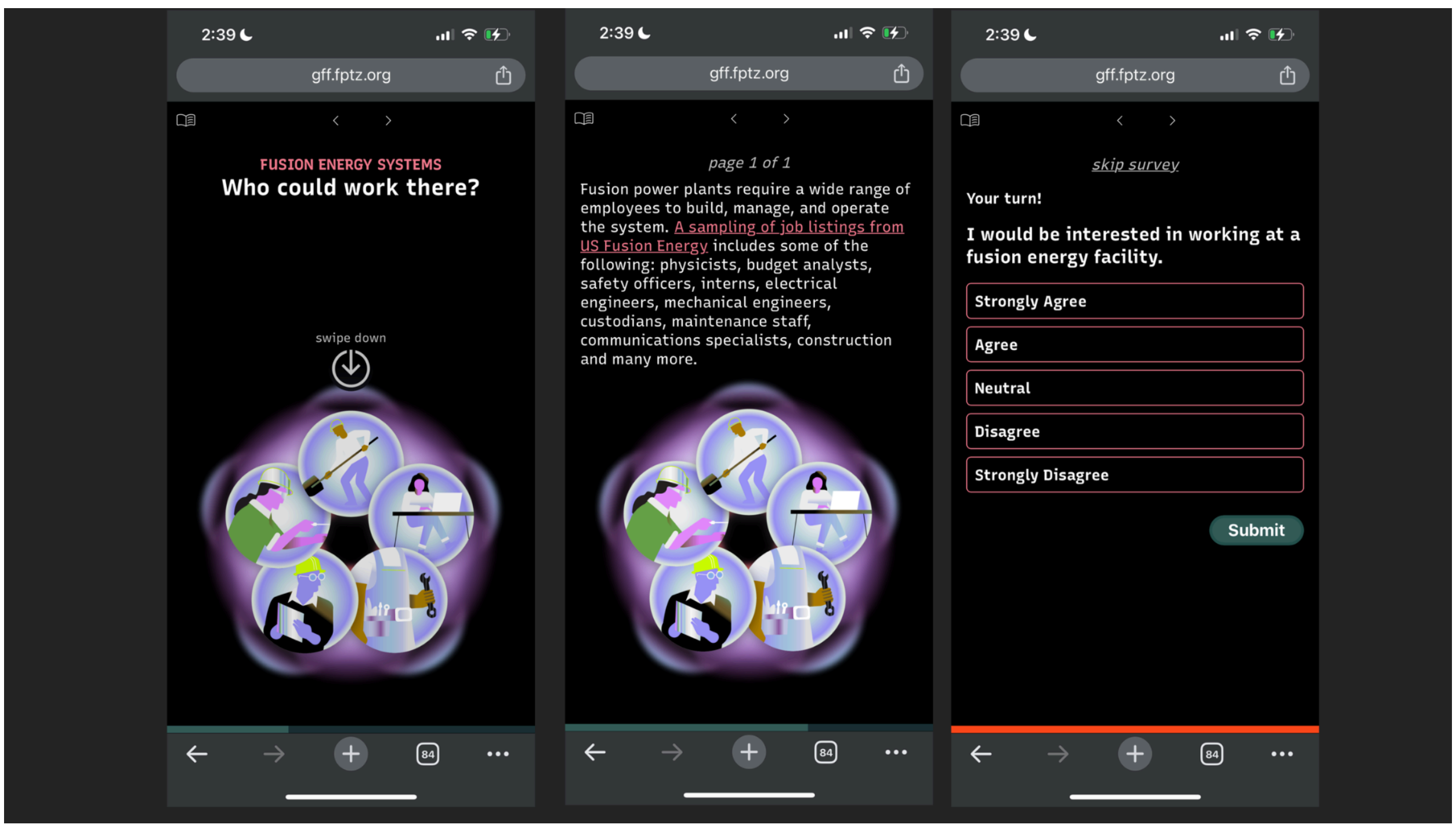

**Figure 11.** The screenshots from Global Fusion Forum above show the text and visuals for the section on "Who could work there" as well as a question users are asked about their interest in working at a fusion facility.

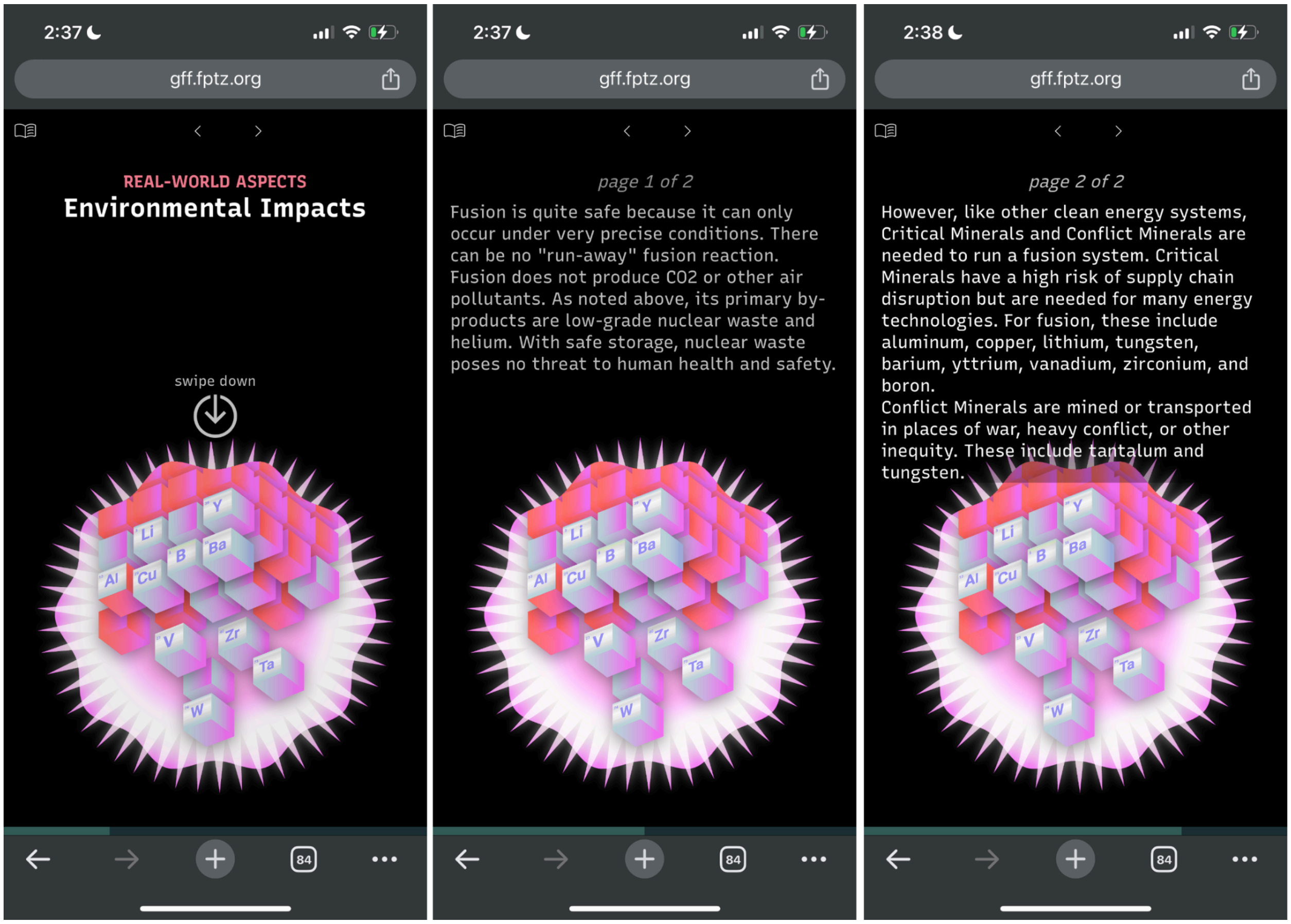

**Figure 12.** An example of the informative text and visuals for the section of environmental impacts of fusion energy

## Speculative design of fusion energy facilities

Following the success of our design workshop with students and community members in the Fall of 2023, we wanted to scale up our participatory approach to designing energy facilities and lower the barriers for participation. Our main motivation for scaling up the approach was to create a community-sourced idea bank of fusion energy facility design concepts from which future fusion energy developers could draw inspiration while designing their own real-world facilities. We initially started with the idea of creating a design-your-own energy facility kit. We imagined this as a physical kit of which we would make hundreds of copies to be sent out to interested individuals across the US, as well as to other countries. In developing this plan, we were confronted with two logistical challenges: the problem of identifying interested users and shipping the kit to them, and then for the users to share their design ideas (either by taking photos or returning the completed kit to us). We also acknowledged that a lack of funding (we were doing this work with our discretionary and internal funds) would be a challenge for scaling up and sustaining this approach over time. While it remains our intention to create a physical kit, we decided, as a starting point, to create an online version of the design-your-own energy facility kit, which lent itself well to creating an online facility design idea bank. Our online design-your-own energy facility platform can be found at [www.imaginaryenergies.com](www.imaginaryenergies.com) . The welcome page and main page for the platform can be seen in Figures 13 and 14.

The website does not require a login and is available free of charge. Users are invited to begin by sharing their most cherished values, which they believe should underpin the development of built environments and systems in the future. On the following page, users are led through a series of simple prompts that ask them to first choose their energy facility type; they can also choose to design a facility made up of multiple energy technologies. For example, a user might design an energy facility that combines fusion energy with solar energy. They are then invited to decide on the size and location of the facility, its aesthetic style, and to describe any key special features and surroundings. Users may also wish to co-locate a waste management facility along with their energy facility. The platform generates a textual description of the facility design based on the values specified by the user on the onboarding page, as well as the facility design choices made on the second page. This prompt can be used to generate an image of a user's hypothetical facility and add their facility design to a publicly available gallery and map. In an upcoming iteration of the platform, users will be able to include a reflection on their facility design (what they like or dislike about it, and what might be missing) and share this reflection along with their facility image in the design gallery. Further, in the current version of the platform, while the users are asked to specify the values, the interpretation of values into design choices is done at least in part by the generative AI model, unless the user specifies how these values should be interpreted as design characteristics through additions to the textual prompt, or through the facility design choices made on the second page. In a future iteration of the platform, we intend to introduce a framework through which users can themselves reflect on their chosen values and interpret them in the context of their facility design.

Initially, the imaginary energies platform was primarily used by our students and colleagues. We have also used imaginary energies in our undergraduate course as well as our summer school program. We have rolled out the platform now to broader publics, and as we start to build the community-sourced idea bank we will periodically analyze and disseminate energy facility design 'trends' we see emerging.
We hope that this easy-to-use, visually appealing platform elicits interest from the public in increasing energy literacy and participating in the discourse and decisions surrounding energy choices for our future.

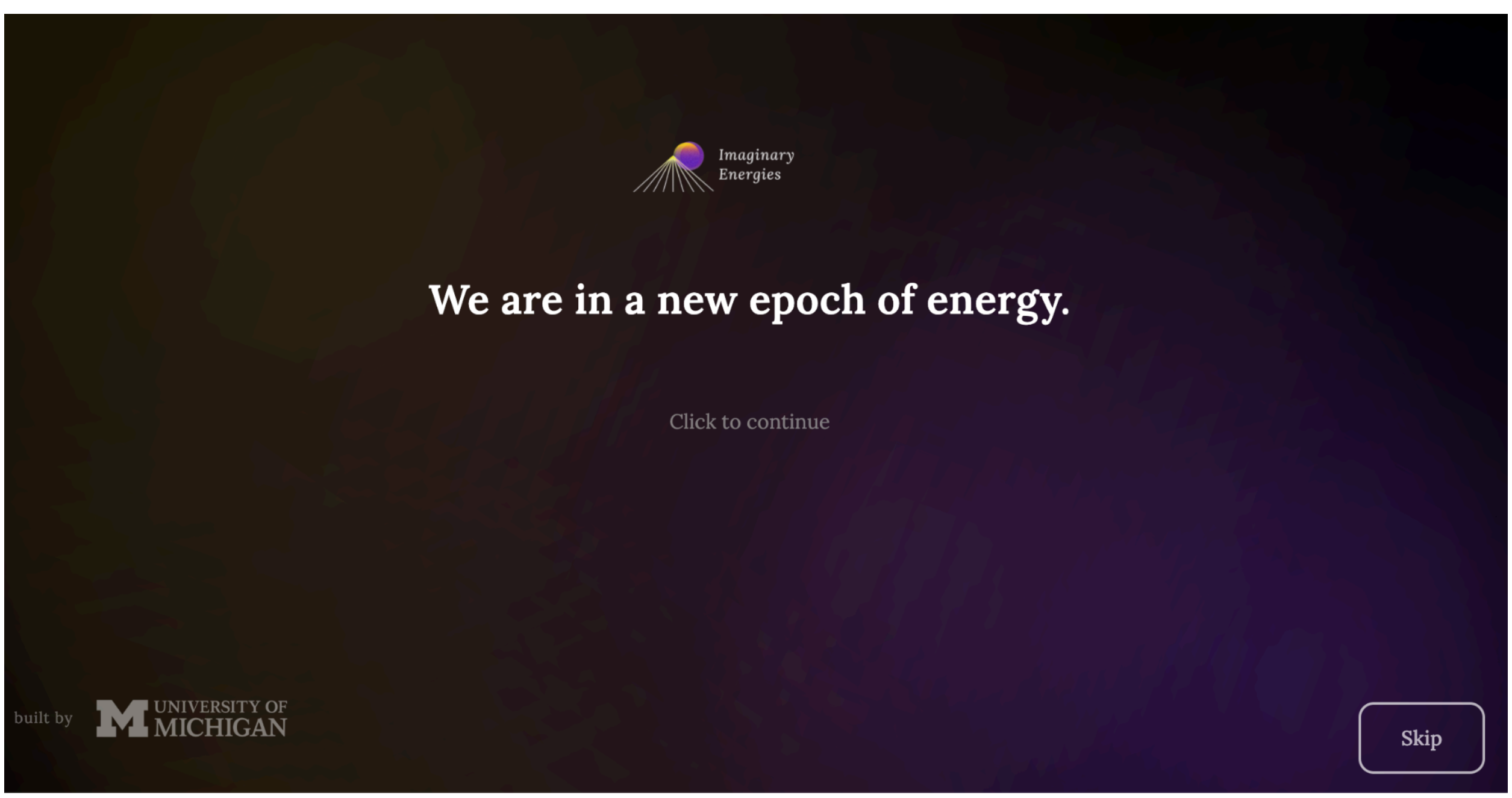

Figure 13: The landing page for Imaginary Energies

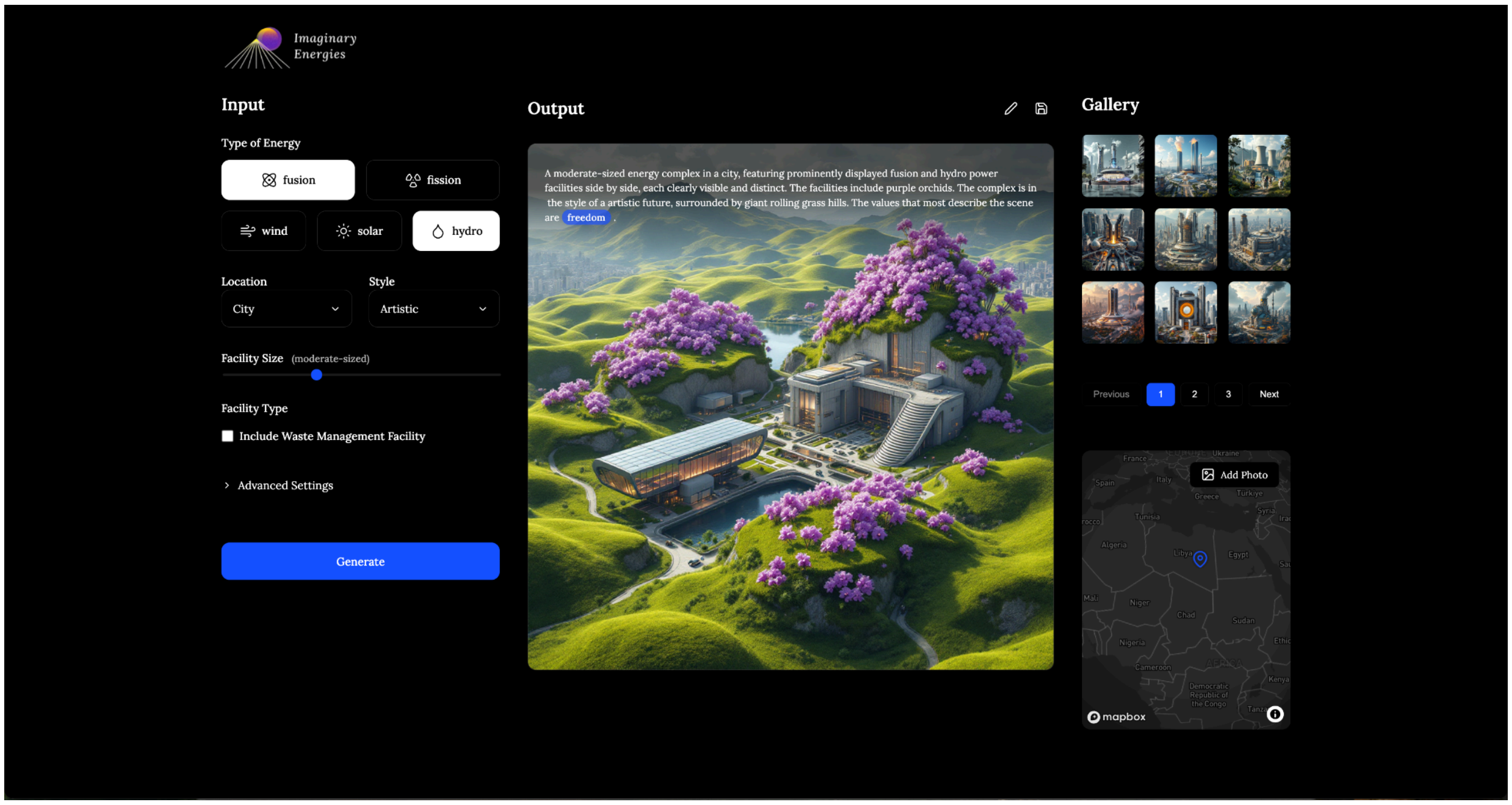

Figure 14: The Design of the Imaginary Energies platform is shown here. Users can specify their facility, type, size, location, as well as select additional energy facility features in the panel on the left. The generated image is seen in the middle panel and the gallery is seen to the right.

## Arts-based engagement

In a further attempt to embody the physicality of fusion and, more broadly, energy-related science and engineering, we have developed “Heartbeat”, a live-streamed interactive website of the plethora of noises experienced in fusion labs and plants. Dubbed so after the creator Morales’ desire to depict “the beating heart of science,” the prototype features the sounds of pumps, magnets, liquids, pipes, and other industrial features of fusion as they are active. Stills from Heartbeat are shown in Figure 15.

The site was coded by Morales so that visual representations of the variance in the sounds fill up the screen (Figure 15), allowing non-technical audiences to viscerally experience the magic of fusion. Underlying the site is a desire to demystify fusion while unveiling the magic details that fascinate scientists and engineers working day in and day out in the field.

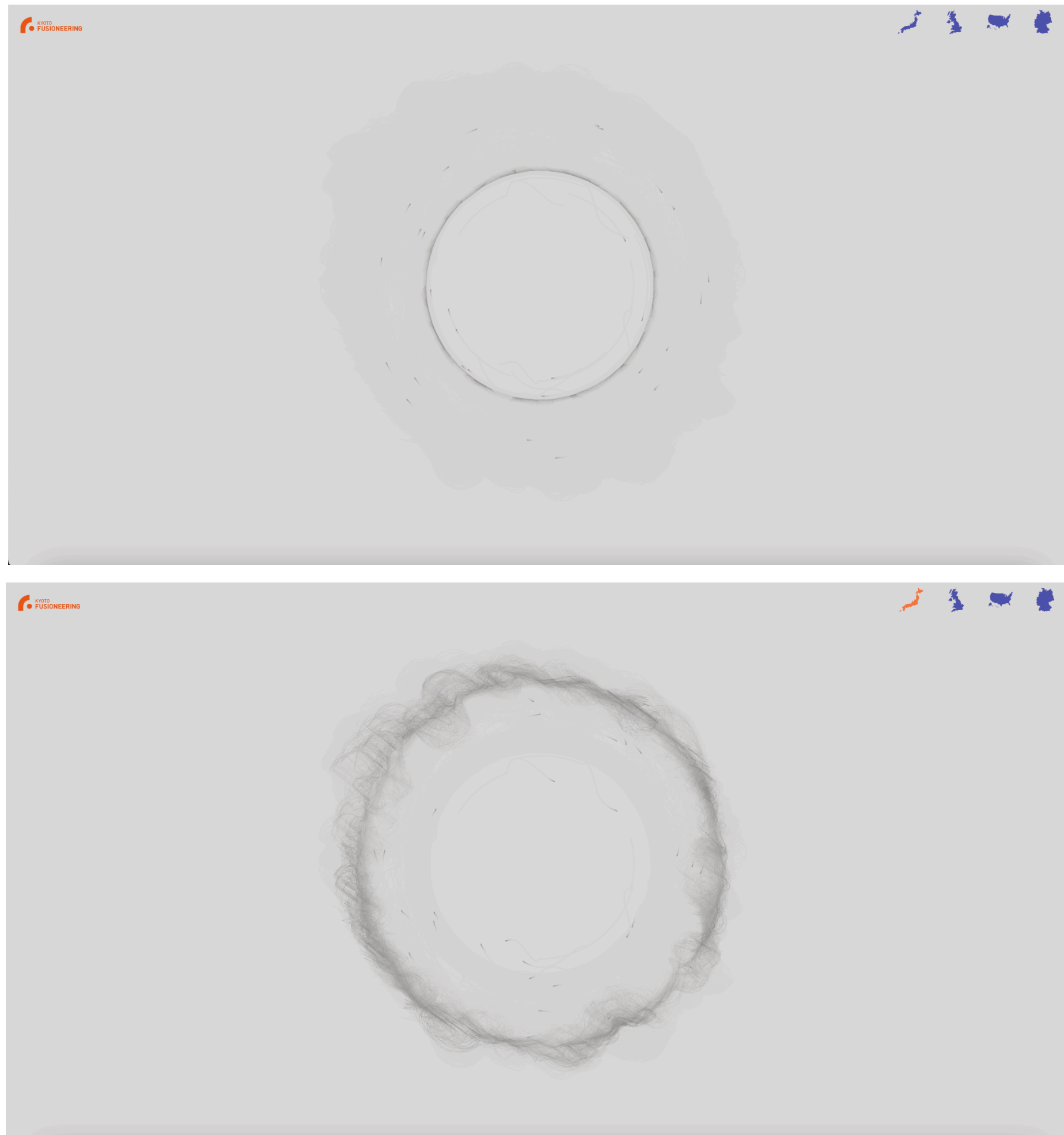

**Figure 15.** This figure shows two stills from Heartbeat. At the top: the site before the user selects a country and a lab to listen to. At the bottom: the turbulence the visualization suffers as the sound of a cryogenic pump is live-streamed to the site.

There is a core theory to Heartbeat: that elevating the mundanity of science to the plane of the aesthetically experienced will lead to higher adoption and curiosity by the broader public writ large. It stands side by side with past exercises in similar vein, like IBM's Pavilion in Expo 67, Disney's "Our Friend The Atom," and the multitude of times NASA participated in World Fairs before and after the first moon landing. In a continuation of this work, we have disseminated a survey to fusion researchers and professionals from around the world to collect the sounds of fusion from their labs and workplaces.

# Toward ecologizing participation

As indicated by the range of case studies presented here, we do not mean to suggest that there is a single 'best' way to invite participation in energy technology related decision-making. Instead, our work has been motivated by creating an ecology of participation, with multiple modalities of engagement to meet varying levels of interest and availability of different individuals. If grown and built out on a larger scale, we believe this approach to engagement and participation can replace the current one-off, instrumental approaches to engagement that have become the norm for large technology or infrastructure projects. Under these one-off or episodic engagement approaches, often motivated by the availability of external funding or the need to answer 'urgent' questions as construed by government or industry, communities and publics are not consulted until the late stages of technology development or project siting, at which point, consultation is essentially a referendum on the project, rather than an opportunity to shape its design and local impacts.

## Cognizance of power dynamics

It is worth noting that while we invite public and community engagement across the case studies described here, each project ultimately was designed predominantly by teams of researchers, and we as researchers, invited community members into research and teaching spaces (for in-person events), paid community members for their time, and where possible, documented our engagement efforts to be able to analyze their outcomes. Although our work has been well-meaning, this imbalance of decision-making power is relevant. Many proponents in the participatory research and practice spheres call for a shift from informing and consulting communities, or involving and collaborating with communities, to empowering them to make decisions. Indeed, the majority of our own efforts can best be described as involving and collaborating with publics and communities. However, as an alternative, we can imagine a society in which communities have their own resources to commission experts to do the kind of work we have done, with decisions on the different modalities of engagement (such as the design of participatory workshops and the online platforms described here) made by communities, with experts deferring to them. Under such arrangements, communities might choose to invite experts into shared public spaces and determine the timing and frequency of engagements to best suit the availability of community members. These questions about collaborating with vs empowering communities raise broader questions about the primacy and politics of expertise. In most societies scientific and technical expertise is ceded to intellectual disciplines and professionals, until controversies or mistrust wrest power from expertise. With a growing international attention, particularly among the youth, to the perils of a changing climate, the need to decarbonize our energy systems, and activism in both of these areas, it is likely that we are moving toward societies in which 'experts' will necessarily need to be open to greater scrutiny. Given the advent of fusion energy in this changing world, it is important for fusion experts to be cognizant of current as well as the changing nature of power, publics, and expertise.

## Mapping our case studies across the values, norms, design space

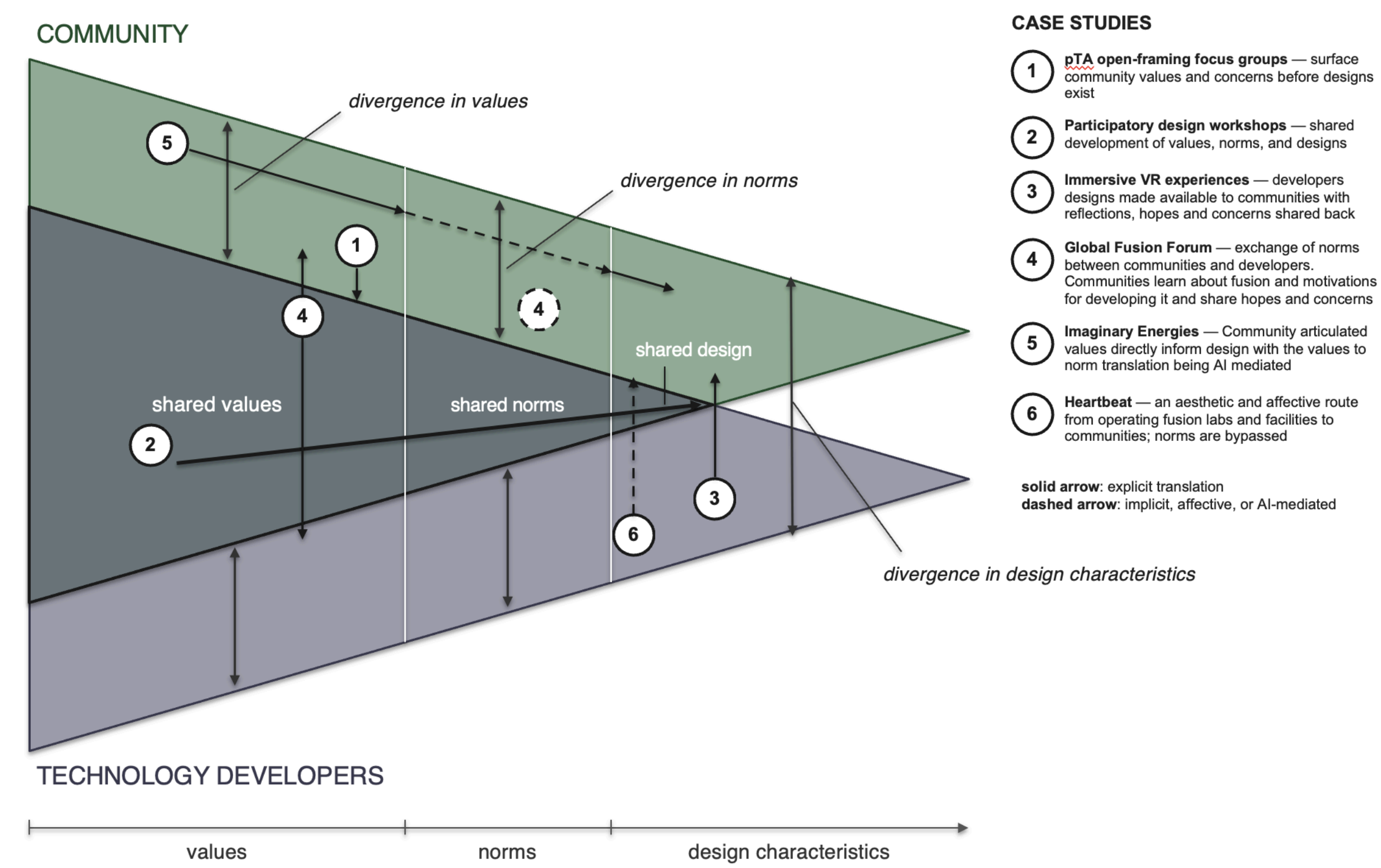

**Figure 16**. Mapping our case studies onto the values, norms, design framework

Each of the case studies inhabits different locations in the values, norms and design characteristics pyramid shown earlier and each has played a different role in interrogating and in some cases, closing, the divergences. The pTA open framing focus groups helped surface community values and concerns prior to the existence of designs. The information yielded by these (as well as potential future) pTA focus groups is valuable because it can support the creation of both norms as well as designs that, from the outset, do not diverge sharply from community values. This information is also valuable because it can signal which technologies or designs are unlikely to succeed in closing the divergences. The immersive VR experiences brought the ITER design to audiences of students as well as community members, enabling them to react to this particular system and share their hopes and concerns. Community input gathered from these immersive experiences of a technology demonstration project such as ITER can inform the development of subsequent fusion energy systems.

Our online platform, Global Fusion Forum, enabled an exchange of norms between community members (users of the platform) and fusion researchers and developers. As users of the platform, community members learned about the fundamentals of fusion energy and the motivations underpinning its development and, as with the VR experiences, were able to share their hopes and concerns – this time with a particular focus on norms (how should fusion systems be developed? Where and how should they be used? What developmental aspects should be prioritized?), rather than designs. The imaginary energies platform enabled community members, again as users, to create their own visualizations of fusion energy facilities. Here, users started by identifying values as well as some design priorities. Where designs were not entirely specified, the translation from values to norms was mediated by the platform itself. This instance is a notable limitation of the platform and future iterations will assist users in arriving at their values to norms translations rather than inserting technology mediated inferences. Heartbeat – which shares the acoustics of fusion facilities with users takes an affective route. The sounds emitted by a fusion laboratory are intended to inspire curiosity, wonder, and a desire to learn more. Notably, this project operates at the design level of the values, norms, design pyramid. Finally, the participatory design workshops occupy a unique role across these case studies because they deliberately traverse the entire pyramid from values, norms to designs, and engage both community members and engineers in interrogating the values, norms, and design divergences and closing them through collaboration and mutual learning.

Ultimately, we view this values, norms, designs framework as an evaluative one that can be used to reflect on the motivations, process, and outcome of participatory engagements ex ante or ex post. We are generally of the view that while engagements that traverse the values, norms, and design pyramid are most valuable, not all engagements (as demonstrated by our own) will be able to do so. Indeed engagements that traverse the entire values, norms, design space are time intensive to create and participate in – for technology developers as well as community members. This is not to say that they should not be undertaken but that they can be completed by other efforts spanning other parts of the values, norms, design space as collective parts of an ecology of participation.

Mapping our efforts across the values, norms, and design space also reveals two notable limitations of our work. The first limitation is that while several of our case studies ask community members to articulate their values or norms, or inspect technology developer's design choices, that dynamic is seldom flipped, with technology developers not being asked to reflect on and share their own values and norms and react to facility preferences and designs that community members may be proposing. This asymmetry is an important opportunity both for future work and for building a more complete ecology of participation. A second shortcoming is that our efforts though seeking to center the environment – whether in the framing of the pTA focus groups, or in the designs created in the participatory design workshops and on imaginary energies – environmental concerns clearly remain secondary to human concerns and the more than human ethic articulated earlier in this paper is only partially satisfied. Future work ought to better explore how to treat the environment as stakeholder in its own right and here the emerging work within the more than human design field and environmental humanities can offer both theoretical and methodological insights.

# A framework for building the ecology of participation

In closing, we would like to offer a framework to fusion researchers and technology developers who are seeking to create their own participatory engagements and thus add to the ecology of participation.

## 1. What part of the divergence are you seeking to learn about or close?

First, consider what part of the divergence – values, norms, or design, you seek to learn or close. Your own role in the fusion research or technology development process might inform which aspect of the divergence is most critical for you to learn about. Those involved in fundamental research and early stage technology design would do well to consider how their values and norms are shaping their research and design choices and whether and how these choices may or may not overlap with those held by publics. These questions cannot be answered in a vacuum or through secondary data and require direct engagement. Values sensitive design or co-design may be valuable methods to incorporate here – again, depending upon what part of the divergence you aim to explore. Perhaps also consider forming collaborations with others in the fusion community motivated by a similar set of questions as a way to build understanding, along with sharable engagement practices and strategies.

## 2. Where is your engagement located across the values, norms, design space?

Next, consider which parts of the values, norms, design space your engagement can traverse. Where does your work begin and where does it end? Are you inadvertently privileging norms over values, design over norms? Can you engage fusion researchers as well as participating community members in a shared articulation of their values or norms or design choices – perhaps only one of these components? Or are there spaces (values, for example) that you cannot immediately engage? In that case, what might you do to close the gap? Design justice literature and methods could help provide answers to some of these questions. Note also the roles of *time* and *trust* in your engagement work. Engagement works on an axis where time allows space for trust to grow. When trust is well-established, and communication lines are strong, project work can move quite quickly (assuming material and regulatory concerns are also addressed).

## 3. Who is missing from your ecology of participation?

A third question to consider is who is missing from a specific engagement or more broadly from the ecology of participation. There may be groups who are underrepresented or those who are indirectly impacted but entirely excluded from the engagement process. Consider ways to bring in those missing collaborators, and whether by digital, in-person, or hybrid means, the mode of communication should be matched to participants' access. The need for high-speed Internet connection or to travel for an in-person meeting may be limiting factors and should be accommodated. Further, as noted above, we identified that in our work, though well-intentioned, the environment was represented at best, indirectly. Careful consideration of representing the environment and its non-human creatures as impacted communities in energy systems should be part of every engagement process. Life-cycle assessment could be useful in this context.

Finally, while it is not possible to engage this group directly, we must consider future generations who will be impacted by current energy system design. Speculative design methods can perhaps help develop these questions and explore their answers, as well as more-than-human design. Initial questions may be, how can we design for those communities who do not exist yet, but will certainly experience net benefits and harms from the systems we create today? As a starting point, it may be worth asking, as noted above, what can we learn from past design work and its enduring impacts, to better care for future generations' energy needs and constraints?

## 4. How can different engagement practices complement one another?

Noting that some engagements will be episodic whereas others will be ongoing, consider how your own engagement relates to other efforts. Where can your engagement begin and where can it leave off in attempting to traverse the space of values, norms, and designs. Over-engagement is perhaps as much a problem as under-engagement. Communities, and individual people, have a heavy slate of daily obligations, so finding a balance where communication lines remain consistent and strong, but community members are not deluged with information and requests for their time is critical. Representation is an important practice, where several dedicated community members can take information back to large groups, and conversely, share group ideas with designers. But this work might well be orchestrated with opportunities for oral deliberation, debate, focus groups, immersive learning experiences, and similar practices as suited to the community interest and access. Participatory design methods can inform much of this work, keeping in mind specific strategies like PTA.

## 5. Which divergences cannot be closed?

As we have noted above, some divergences will ultimately be intractable. They can be interrogated and understood but not closed. We urge fusion researchers and developers not to force closure of these divergences and secure acceptance for a technology when such acceptance cannot be won. Indeed, as we noted earlier, pursuit of such acceptance has created antagonistic relationships between experts and societies. The path to energy system security is unlikely to mean all communities will accept or even need the same type of system or source of power. For those that choose to go another direction, it may be useful to see that choice as a kind of healthy diversification rather than a loss or lack. Again, speculative design may be valuable here, to help articulate desirable futures in the context of what is possible, plausible, and probable.

# Conclusion

In closing, we want to reiterate the wide range of expertise and knowledge that made the work reported here possible. Forms of expertise held by our team span technical communication, participatory design, plasma physics, nuclear engineering, scientific illustration, UX design, web-application development, design research, public policy, computer science, and 3D modeling and visualization. As is evident, many forms of expertise outside plasma physics and fusion technology have been critical in this work. On a larger scale, we hope this signals to the readership of this journal that building an ecology of participation is also an exercise in forging collaborations across multiple disciplines and forms of expertise. The successful path forward for fusion energy system development may rely on this kind of multidisciplinary collaboration.

# Author declarations

## Conflict of interest

The authors have no conflicts to disclose.

## Ethics approval

The participatory design workshops and the student reflections on the virtual reality models described in this paper were conducted under a protocol reviewed by the University of Michigan Institutional Review Board and determined to be exempt (HUM00240439, "Integrating Socially Engaged Design of Nuclear Energy Technologies in ENGR100," approved October 30, 2023). All participants provided informed consent prior to participation.

The participatory technology assessment focus groups were conducted under a protocol approved by the University of Wisconsin–Madison Institutional Review Board (IRB submission ID: 2023-1057). Participants were 18 or older and, prior to participation, provided informed consent to participate and to audio and video recording. Participants received a stipend for completing both sessions, and all survey, worksheet, and discussion data were collected confidentially and reported only in aggregate or anonymized form.

## Data availability statement

The data that support the findings of this study are available on request from the corresponding author. The data are not publicly available due to privacy and ethical restrictions: the focus group and workshop data include recordings, worksheets, student reflections, and discussion notes collected under informed consent agreements that limit their use to the research team and sharing to aggregate or anonymized form.